\documentclass[preprint,aps,12pt,showpacs,nofootinbib,tightenlines]{revtex4}
\usepackage{amsmath}
\usepackage{amssymb}
\usepackage{epsfig}
\usepackage{graphicx}
\usepackage{ctex}

\begin{document}
\def\pslash{\rlap{\hspace{0.02cm}/}{p}}
\def\eslash{\rlap{\hspace{0.02cm}/}{e}}
\title {The signatures of the quintuplet leptons at the LHC }
\author{%
You Yu  $^{1}$}
 \author{ Chong-Xing Yue $^{1}$}\email{cxyue@lnnu.edu.cn}
\author{ Shuo  Yang $^{2}$ }\email{yangshuo@dlu.edu.cn}

\address{%
$^1$ Department of Physics, Liaoning Normal University, Dalian 116029,  China\\
$^2$ Physics Department, Dalian University, Dalian 116622,  China
 \vspace*{1.5cm}}

\begin{abstract}
We investigate the production and detection prospects for the quintuplet heavy leptons at the LHC
in the context of a new model which is proposed as a viable and testable solution to the neutrino
 mass problem. We classify the signals, carry out a full simulation on the signals and the relevant
 backgrounds at the 14 TeV LHC. After applying suitable kinematic cuts, the background events are
 substantially suppressed. The signals of the heavy leptons might be detected at the 14 TeV LHC.
\end{abstract}
 \maketitle

\newpage
\noindent{\bf 1. INTRODUCTION }

\vspace{0.5cm}The Standard Model (SM) of particle physics has successfully described experimental data so far.
The Large Hadron Collider (LHC) discovered a SM-like Higgs particle \cite{Higgs} with mass around
125 GeV on July 4, 2012, which might be treated as significant evidence for further proving the SM.
However, the SM still has theoretical shortcomings, like small neutrino masses. Many new physics models
beyond the SM have been proposed aiming to solve this problem.
Three types of seesaw mechanisms can explain the small neutrino masses
 by introducing extra particles at a high scale, which generates the neutrino masses through the effective
 dimension-five Weinberg operator LLHH \cite{LLHH} at tree level. The extra particles correspond to a
 heavy fermion singlet in type I, a scalar triplet in type II and a fermion triplet in type III, respectively
 \cite{one,two,three}. Other mechanisms can also account for the small neutrino masses and should be explored.



In addition to the canonical seesaw mechanisms, the cascade seesaw mechanism \cite{liaoyi}
was proposed to generate neutrino masses through a higher dimension ($5+4n$) operator.
Similar ideas for generating the neutrino masses via the higher dimensional operators are
considered in Refs. \cite{09052710,0105281}.
The case $n=1$ \cite{12046599} corresponds to the minimal version of the cascade seesaw, which will be considered in this paper. In addition to SM particles, this model introduces three generations Majorana quintuplets $\Sigma_R$ with zero hypercharge transforming as (1,5,0) under the SM gauge group $SU(3)_C \times SU(2)_L \times U(1)_Y$ and a scalar quadruplet $\Phi$ transforming as (1,4,-1).
In this model, small neutrino mass $m_\nu \sim  \frac{v^6}{\mu_{\Phi}^4 M_k}$ is obtained via an effective dimension-nine operator (LLHH)(H$^\dag$H)$^2$.
Here, $v$ is the vacuum expectation value (vev) of the SM Higgs, $\mu_{\Phi}$ is the mass scale of the scalar quadruplet and $M_k$ is the mass
scale of the $k$ generation fermion quintuplet.\footnote {Assuming $\mu_{\Phi}\sim M_k \sim M$, the neutrino mass is approximately $m_\nu \sim  \frac{v^6}{M^5}$.}
This is different from the conventional three types of seesaw formula $m_\nu \sim v^2/M$,
where $M$ is the scale of the new physics. In this model, neutrino masses can also be generated by a radiative diagram which induce
a dimension-five operator with additional loop suppression. This loop mass of neutrino is not achieved in the type III seesaw model.
This new model with Majorana quintuplets is therefore something of a hybrid between the traditional seesaw
mechanisms and the traditional radiative models of neutrino masses.

The fermion quintuplet contains the doubly charged, singly charged and neutral heavy leptons.
The doubly charged heavy leptons are salient feature appearing in many models,
which can provide two same-sign leptons as the smoking gun for the
scenarios \cite{sourceLNV}. Any signal for such kind of new leptons in future high energy experiments
will play an important role in testing the SM flavor structure and discovery of the new physics.
Many studies have been carried out on single production and pair production of the doubly charged lepton \cite{hantao, yuyou, xiayun, 13071711, many}.
In addition, Refs. \cite{zheng, mateng13, mateng14, dingran} have also studied the phenomenology
of doubly charged heavy leptons, singly charged heavy leptons and neutral leptons in exotic lepton multiplet models. The doubly charged fermions also appear in flavor models in warped extra dimensions and in some general models \cite{shengao1, shengao2, mod1, mod2}.
In this paper, we calculate the production of the doubly charged, singly charged and neutral heavy leptons,
and analyze the signals and backgrounds at the LHC in the context of this zero hypercharge quintuplet fermion model.

The heavy leptons have been searched at the LHC, which has already posed significant bounds on the masses of
these exotic leptons. But such searches depend strongly on the flavor structure.
 Light states are still allowed if their couplings are suppressed, if they decay into final states affected by large backgrounds, or if they
are not efficiently produced at the LHC.
Reference \cite{hantaomass} has given that the current lower bound for
the mass of a generic charged lepton is 100.8 GeV.
The ATLAS and CMS collaborations have recently provided lower bounds on the mass of long lived multi-charged
particles, which decay outside the detector \cite{longlivedmass1, longlivedmass2}. These constraints don't
apply to promptly decaying particles like those that we consider here.
The stronger bounds on the masses of the exotic leptons are provided from the
generic searches for lepton-rich final states at 8 TeV \cite{ATLAS, CMS}.
These constraints will be considered when we discuss the mass range of the signals in the following analysis.

The rest of the paper is organized as follows. In Sec.II, we review the basic content of the new model.
In Sec.III, we calculate the cross sections of the heavy leptons and
present the phenomenological analysis for several interesting search channels.
Our main results are recapitulated in Sec.IV.

\vspace{0.5cm} \noindent{\bf 2. THE MODEL WITH MAJORANA QUINTUPLETS}

Recently, Ref. \cite{12046599} proposed an alternative model to address neutrino mass problem.
This model is based on the SM gauge symmetry $SU(3)_C\otimes SU(2)_L \otimes U(1)_Y$.
In addition to the SM particles, three generations of Majorana quintuplets
$\Sigma_R=(\Sigma_R^{++},\Sigma_R^+,\Sigma_R^0,\Sigma_R^-,\Sigma_R^{--})$ with zero hypercharge are introduced
transforming as (1,5,0) under the SM gauge group, where $R$ denotes the chirality.
The fermionic quintuplet with zero hypercharge is treated as the simplest generation
of type III seesaw Majorana triplet to a higher isospin multiplet.
It can also provide a viable minimal dark matter candidate \cite{dark}.
In addition to the SM Higgs doublet $H=(H^+,H^0)$, a scalar quadruplet
$\Phi=(\Phi^+,\Phi^0,\Phi^-,\Phi^{--}) \sim (1,4,-1)$ is introduced, its neutral member $\Phi^0$ acquires a
 nonvanishing vev and generates neutrino masses.
  The masses of the new particles predicted in this model are naturally at the TeV scale.


The gauge invariant and renormalizable Lagrangian involving $\Sigma_R$ and $\Phi$ can be given as \cite{12046599}:
\begin{eqnarray}
\mathcal{L}=\overline{\Sigma_R}i \gamma^\mu D_\mu \Sigma_R+(D^\mu \Phi)^\dag(D_\mu \Phi)-(\overline{L_L}Y\Phi\Sigma_R+
\frac{1}{2}\overline{(\Sigma_R)^C} M\Sigma_R+H.c.)-V(H,\Phi),
\end{eqnarray}
where the $\overline{L_L}Y\Phi\Sigma_R$ term is the Yukawa coupling among the scalar quadruplet, the SM left-hand
 lepton doublet and the fermion quintuplet, and $Y$ is the Yukawa-coupling matrix.
 The $\frac{1}{2}\overline{(\Sigma_R)^C} M\Sigma_R$ term is the Majorana mass term of the fermion quintuplet,
 $M$ is the mass matrix of the heavy leptons. The $V(H,\Phi)$ term is the scalar potential, whose expression
 can be given as follows \cite{12046599}:

\begin{eqnarray}
V(H,\Phi)&&=-\mu_H^2H^\dag H+\mu_\Phi^2 \Phi^\dag \Phi+\lambda_1(H^\dag H)^2+\lambda_2 H^\dag H \Phi^\dag \Phi + \lambda_3 H^*H\Phi^*\Phi
\nonumber \\
&&+(\lambda_4 H^*HH\Phi+H.c.)+(\lambda_5 HH\Phi \Phi+H.c.)+(\lambda_6 H\Phi^* \Phi\Phi+H.c.)
\nonumber \\
&&+\lambda_7(\Phi^\dag \Phi)^2+\lambda_8 \Phi^* \Phi \Phi^*\Phi.
\end{eqnarray}

The neutral components of the Higgs doublet and scalar quadruplet $H^0$,
$\Phi^0$ are all responsible for the electroweak symmetry breaking. The vev's of these neutral scalar fields
are  $v$ and $v_\Phi$. There are
\begin{eqnarray}
v=174{\rm GeV},  ~~~~~~~     v_\Phi\simeq -\frac{1}{\sqrt{3}}\lambda_4^*\frac{v^3}{\mu_\Phi^2}.
\end{eqnarray}
$v_\Phi$ will contribute the mass corrections of $M_Z$ and $M_W$, it will confront the constraints from $\rho$ parameter.
The contribution of the model to $\rho$ parameter expression is $6 v_\Phi^2/v^2$. We take the $\rho$ parameter
measurement $\rho=1.0004^{+0.0003}_{-0.0004}$~\cite{electroweakparameter} reported by the Particle Data Group to get $v_\Phi$ limit, $v_\Phi\lesssim 1.9$ GeV.

The Majorana mass matrix of the light neutrino induced from diagonalizing the neutral lepton masses is given by
\begin{eqnarray}
m_\nu^{tree}=-\frac{1}{2} v_\Phi^2 Y M^{-1} Y^T.
\end{eqnarray}
In the basis where the matrix of heavy leptons is real and diagonal, $M=diag(M_1,M_2,M_3)$, and
utilizing the expression in Eq.(3), we can get the tree-level neutrino mass expression which corresponds to
the dimension-nine seesaw operator
\begin{eqnarray}
(m_\nu)_{ij}^{tree}=-\frac{1}{6}(\lambda_4^*)^2 \frac{v^6}{\mu_\Phi^4} \sum \limits_{k} \frac{Y_{ik} Y_{jk}}{M_k}.
\end{eqnarray}

In addition to the tree-level neutrino mass, the neutrino mass can also be generated at one-loop level,
which has the following form
\begin{eqnarray}
(m_\nu)_{ij}^{loop}=-\frac{5 }{24 } \frac{\lambda_5^* v^2}{\pi^2}\sum \limits_{k} \frac{Y_{ik} Y_{jk}M_k}{m_\Phi^2-M_k^2}[1-\frac{M_k^2}{m_\Phi^2-M_k^2} ln\frac{m_\Phi^2}{M_k^2}].
\end{eqnarray}
In the case of $m_\Phi^2 \simeq M_k^2$, the neutrino mass induced at one loop can be approximately expressed as
\begin{eqnarray}
(m_\nu)_{ij}^{loop}=-\frac{5}{48} \frac{\lambda_5^* v^2}{\pi^2} \sum \limits_{k} \frac{Y_{ik} Y_{jk}}{M_k}.
\label{neutrinomass}
\end{eqnarray}
If we consider the tree-level and loop-level contributions together, the neutrino mass is given by
\begin{eqnarray}
(m_\nu)_{ij}&&=(m_\nu)_{ij}^{tree}+(m_\nu)_{ij}^{loop}
\nonumber \\
&&=-\frac{1}{6}(\lambda_4^*)^2 \frac{v^6}{\mu_\Phi^4} \sum \limits_{k} \frac{Y_{ik} Y_{jk}}{M_k}+\frac{-5}{24} \frac{\lambda_5^* v^2}{\pi^2} \sum \limits_{k} \frac{Y_{ik} Y_{jk}M_k}{m_\Phi^2-M_k^2}[1-\frac{M_k^2}{m_\Phi^2-M_k^2} ln\frac{m_\Phi^2}{M_k^2}] .
\end{eqnarray}

Both the heavy leptons ($\Sigma^0$, $\Sigma^\pm$, $\Sigma^{\pm\pm}$) and the scalars ($\Phi^0$, $\Phi^\pm$,$\Phi^{--}$)
 can couple to the SM particles in the new model.
However, we don't consider the phenomenology of the quadruplet scalars in this paper.
We only give the Feynman rules of the heavy leptons to the SM particles, which are related to
 our calculation, can be written as

\begin{eqnarray}
&&\Sigma^0 l W: -\frac{e}{S_W} V_{l\Sigma} \gamma^\mu P_L ,
\nonumber \\
&&\Sigma^0 \nu Z: \frac{1}{2\sqrt{2}} \frac{e}{S_W C_W}[V^\dag_{PMNS} V_{l\Sigma} \gamma^\mu P_L- V^T_{PMNS} V_{l\Sigma}^*\gamma^\mu P_R],
\nonumber \\
&&\Sigma^+ l Z: \frac{\sqrt{3}}{4}\frac{e}{S_W C_W} V_{l\Sigma}^* \gamma^\mu P_R,
\nonumber \\
&&\Sigma^{++} l W: \sqrt{\frac{3}{2}}\frac{e}{S_W} V_{l\Sigma}^*  \gamma^\mu  P_R,
\end{eqnarray}
where $S_W=sin\theta_W$, $C_W=cos\theta_W$, $\theta_W$ is the Weinberg angle, and $V_{PMNS}$ is the $3 \times 3$
Pontecorvo-Maki-Nakagata-Saki (PMNS) matrix \cite{PMNS}, $P_L(P_R)$ is the left-hand (right-hand)
projection operator. $V_{l\Sigma}$ describes the mixing of the heavy leptons and the SM leptons,
its expression is given by
\begin{eqnarray}
V_{l\Sigma}=(v_\Phi Y M^{-1})_{l\Sigma}.
\end{eqnarray}
This variable is proportional to $\sqrt{m_\nu/M_\Sigma}$ which can reach $10^{-7}$ when we take neutrino mass
$m_\nu\sim 0.1$ eV \cite{electroweakparameter} and set the parameters as $Y \sim 10^{-3}$, $\lambda_4 \sim 10^{-2}$ and $\lambda_5 \sim 10^{-4}$ \cite{12046599}.

\vspace{0.5cm} \noindent{\bf 3. PHENOMENOLOGY OF THE QUINTUPLET LEPTONS AT THE LHC}

 \vspace{0.5cm} In this section, we will discuss the phenomenology of the heavy leptons at the LHC.
The productions of the heavy leptons are dominated via the Drell-Yan process mediated by the SM gauge bosons
$\gamma$, $Z$ and $W$. The effective cross sections $\sigma(s)$ can be evaluated from $\hat{\sigma}(\hat{s})$ by
convoluting with $f_{q_1/p}(x_1)$ and $f_{q_2/p}(x_2)$,
\begin{eqnarray}
\sigma(s)=\int^1_{x_{min}} dx_1 \int^1_{x_{min}/x_1} dx_2 f_{q_1/p}(x_1) f_{q_2/p}(x_2) \hat{\sigma}(\hat{s}),
\end{eqnarray}
where $\hat{s}=x_1x_2s$ is the effective center-of-mass (c. m.) energy squareD for the partonic process,
and $x_{min}=4 M_\Sigma^2/s$.
 For the quark distribution functions $f_{q_1/p}(x_1)$ and $f_{q_2/p}(x_2)$, we will use the form
given by the leading order parton distribution function CTEQ6L1 \cite{CTEQ}.
The cross sections have been calculated using tree-level matrix elements generated by MadGraph package \cite{MadGraph}.
The SM parameters are taken as
$M_W=80.4$ GeV, $M_Z=91.2$ GeV and $S_W^2=0.231$ \cite{parameter}.
In Fig.1, we show the production cross sections versus the heavy lepton
 mass $M_\Sigma$ at the 8 (14) TeV LHC.
It is obvious that all the cross sections decrease with the increase of the heavy lepton mass $M_\Sigma$.
 The cross section of $\Sigma^{++}\Sigma^{--}$ production is the largest which can reach 2976 fb for $M_\Sigma=300$ GeV
 and the c. m. energy $\sqrt{s}=$14 TeV.
The $\Sigma^+ \Sigma^-$ production has the smallest cross section, for 200 GeV$\leq$ M$_\Sigma \leq$1000 GeV,
its value is in the range of $744$ fb $\sim 0.6$ fb at the 14 TeV LHC.
 In the following, we will focus on their signals and backgrounds at the LHC.

\begin{figure}[htb]
\vspace{-0.5cm}
\begin{center}
 \epsfig{file=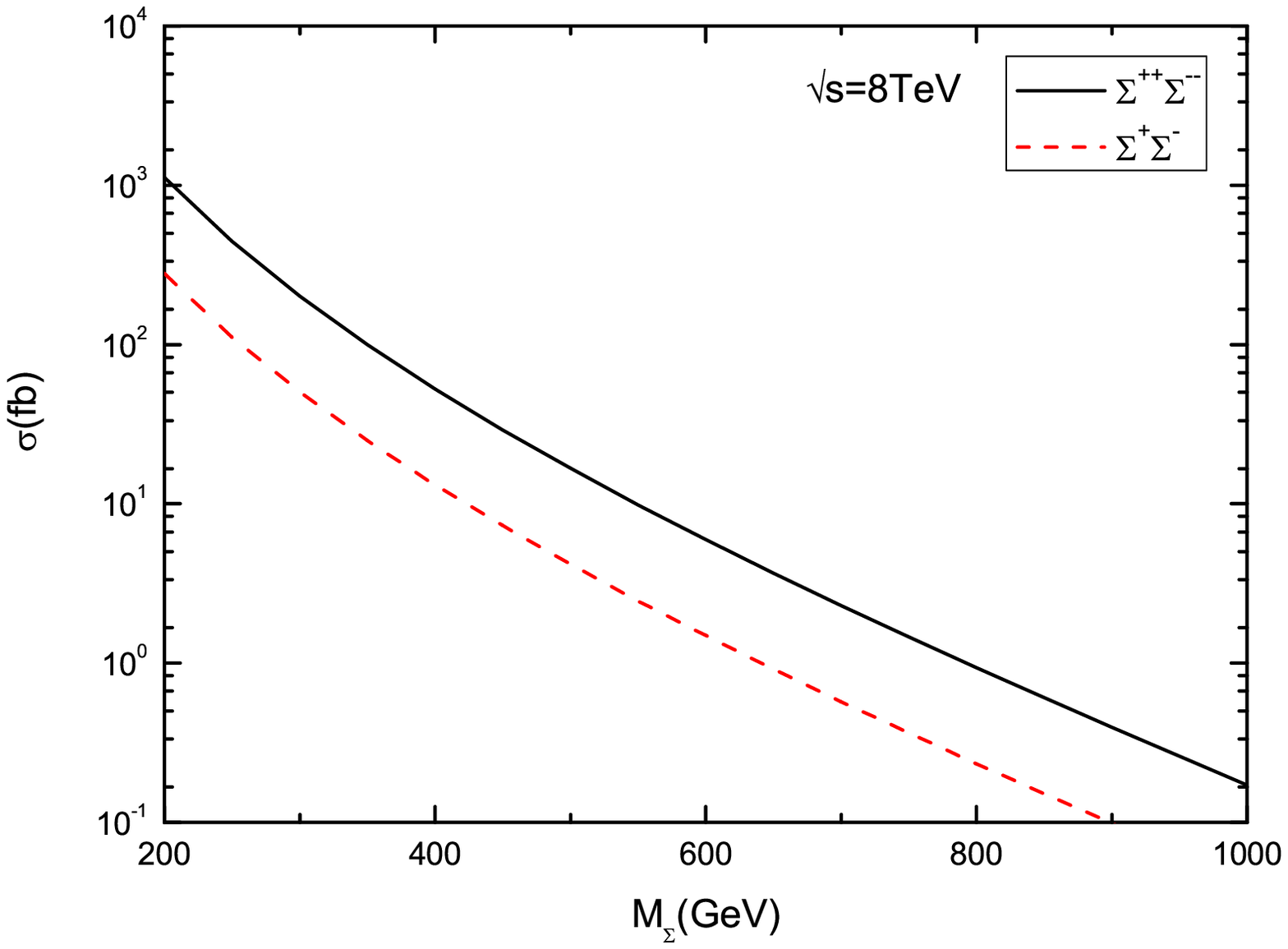,width=220pt,height=180pt}
  \epsfig{file=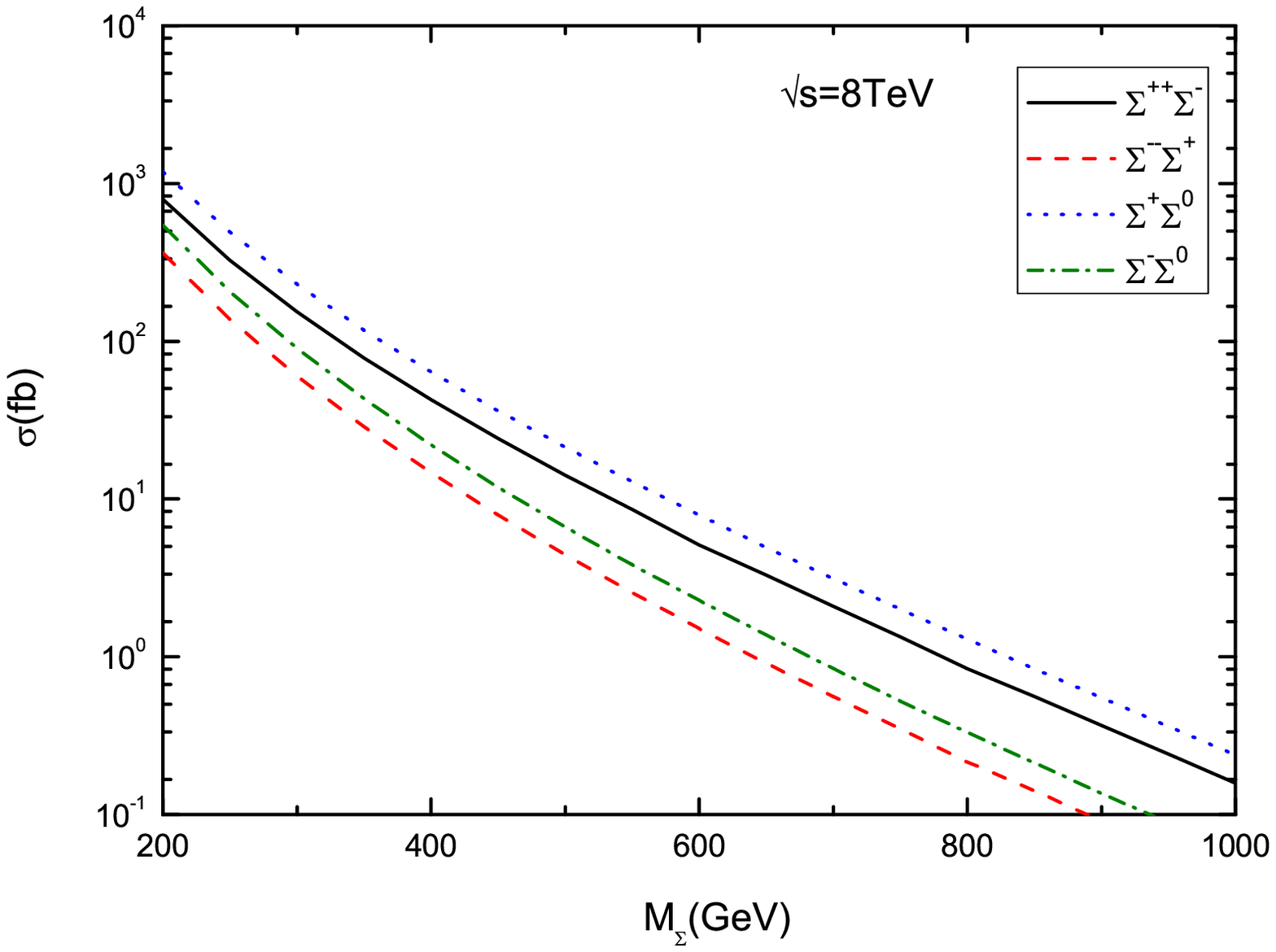,width=220pt,height=180pt} \hspace{-0.5cm}
\nonumber \\
  \epsfig{file=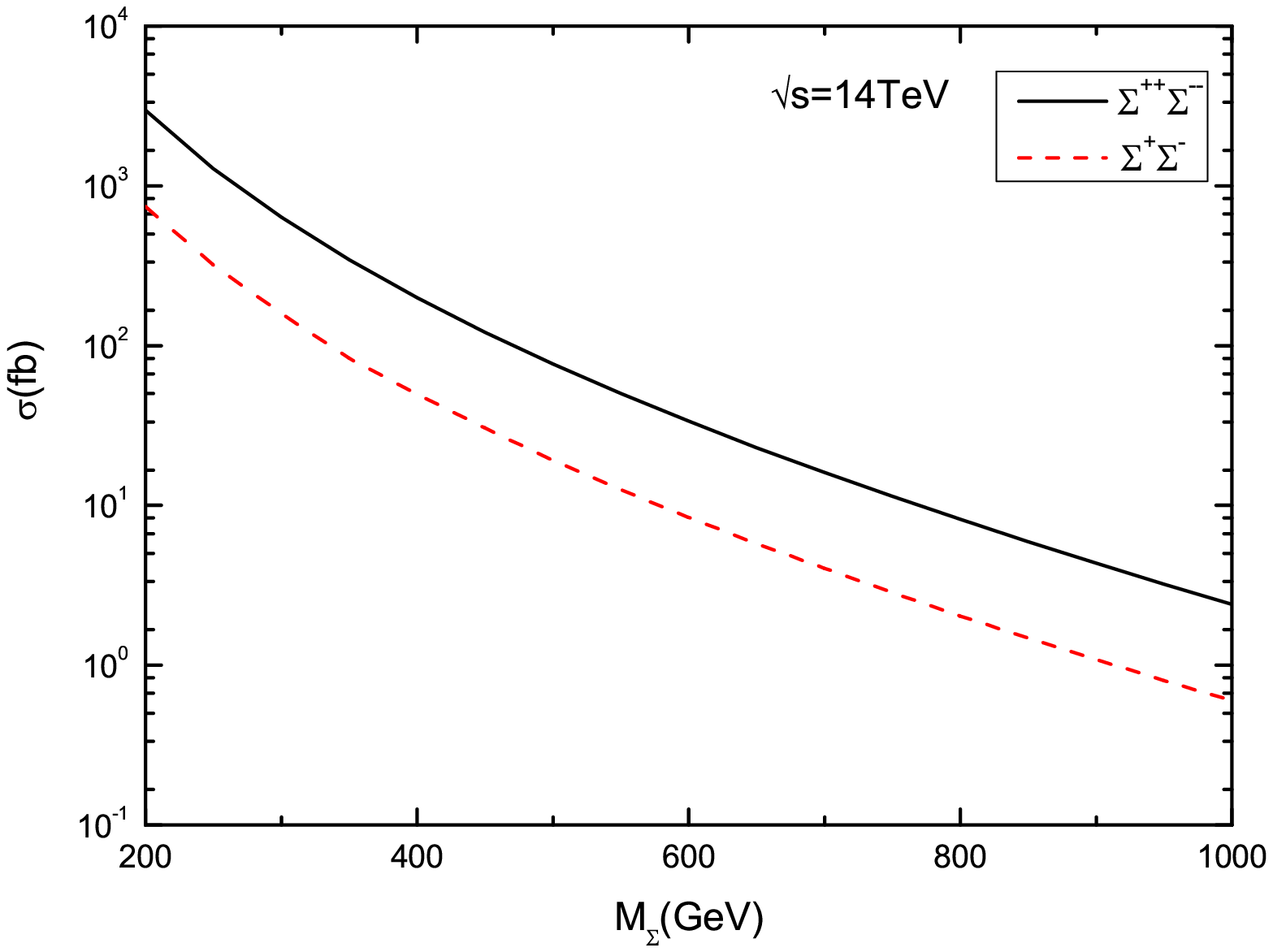,width=220pt,height=180pt}
   \epsfig{file=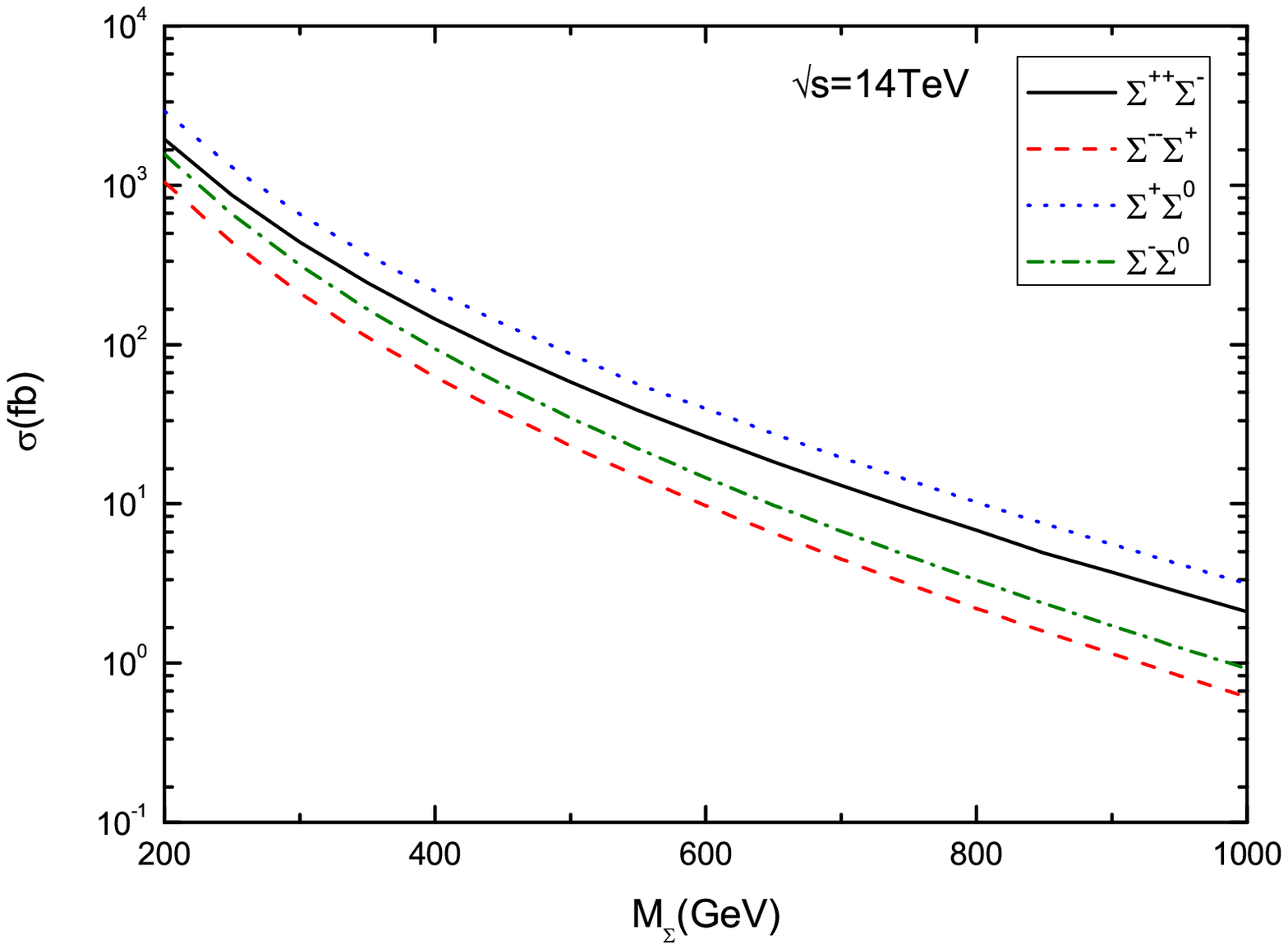,width=220pt,height=180pt} \hspace{-0.5cm}
 \hspace{10cm}\vspace{-1cm}
 \vspace{0.5cm}
\caption{The cross sections of the heavy lepton pair or associated productions
 as a function  \hspace*{1.3cm} of the mass $M_\Sigma$
for the c.m. energy $\sqrt{s}$=8 TeV and 14 TeV.} \label{ee}
\end{center}
\end{figure}

To discuss  the signatures of the heavy quintuplet leptons, one needs to understand their decay properties
to the SM particles. For the most characteristic particle in the model, the doubly charged heavy lepton
$\Sigma^{\pm\pm}$ can only decay to a SM charged lepton with a same-sign $W$ boson $\Sigma^{\pm\pm}\rightarrow l^\pm W^\pm$.
As for $\Sigma^0$, it can decay to $l^\pm W^\mp$ and $\nu Z$. And $\Sigma^\pm$ can decay to $l^\pm Z$
and $\nu W^\pm$. The decay widths sum over the three generations of leptons. The detailed formulas for all of these decay channels are listed in Ref. \cite{12046599}. There are small mass differences between two components of $\Sigma$ quintuplet induced by loops of SM gauge bosons, which are far smaller than the mass scale of $\Sigma$. For $M_{\Sigma}=$400 GeV, the mass differences are $ M_{\Sigma^{++}} - M_{\Sigma^{+}} \simeq 490$ MeV and $ M_{\Sigma^{+}} - M_{\Sigma^{0}} \simeq 163$ MeV, this will induce additional decay channel, such as $\Sigma^i \rightarrow  \Sigma^j\pi^{+}$. However, these decays are suppressed by narrow phase space.
Thus, we take $M_{\Sigma^{++}}\approx M_{\Sigma^{+}} \approx M_{\Sigma^{0}}$ in the following.
The branching widths and the total width of the heavy leptons are proportional to the square of the mixing
 matrix $|V_{l \Sigma}|^2$. In addition, $V_{l \Sigma}$ affects the reconstructed distribution of the signal events.
From the experimental point of view, the mixing matrix $V_{l\Sigma}$ decides the contributions to the
lepton flavor violating (LFV) processes. Thus, the experimental upper bounds on the branching ratios (BRs) of the radiative
LFV decays, for instance, BR$(\mu \rightarrow e \gamma)<5.7 \times 10^{-13}$ \cite{LFV1} and
 BR$(\mu \rightarrow 3e)<1.0 \times 10^{-12}$ \cite{LFV2} can give constraints on $V_{l\Sigma}$.
 We take typical value $V_{l\Sigma}=3.5\times 10^{-7}$ in this paper. Due to the multiple decay modes of the heavy leptons and the SM gauge bosons, we classify the signals in terms of the charged lepton multiplicity as following. And then we consider two typical
cases, $M_\Sigma$=300 GeV and 500 GeV, to perform a full simulation at the 14 TeV LHC.

In order to simulate the unweighted events more realistically at the parton level, we smear the energies of the final
state lepton and jets according to the assumption of the Gaussian
resolution parametrization
\begin{eqnarray}
\frac{\delta(E)}{E}=\frac{a}{\sqrt{E}}\oplus b,
\end{eqnarray}
where $\frac{\delta(E)}{E}$ is the energy resolution, $a$ is a sampling term, $b$ is a constant term,
and $\oplus$ denotes a sum in quadrature. We take $a=5\%$, $b=0.55\%$ for leptons and $a=100\%$, $b=5\%$
 for jets \cite{smear}.

 The following basic selection cuts are applied to all of the signal and background events while generating
 events in MadGraph,
\begin{eqnarray}
&&p^l_T>15{\rm GeV}, ~~~~~~~~~|\eta_l|<2.5,~~~~~~~\rlap/E_T > 25{\rm GeV}
 \nonumber \\
&&p^j_T>20{\rm GeV}, ~~~~~~~~|\eta_j|<2.5,
 \nonumber \\
&&\Delta R_{ll}>0.3, ~~~~~~~~~\Delta R_{jl}>0.4,~~~~~~~~\Delta R_{jj}>0.4,
\end{eqnarray}
where $p_T$ denotes the transverse momentum, $\rlap/E_T$ is the missing transverse momentum from the invisible
neutrino in the final states,
$\Delta R_{ij}$ is defined as $\Delta R_{ij}=\sqrt{(\Delta \eta_{i,j})^2+(\Delta \phi_{i,j})^2}$,
where $\Delta\eta$ is
the rapidity gap and $\Delta\phi$ is the azimuthal angle gap between the particle pair ($i, j= l, j$).
 For the SM leptons, we only consider an electron and a muon in signal simulation and take the
 lepton-tagging efficiency $\epsilon_l=90\%$. The light jet $j$ means light quarks or gluons.
 After the basic cuts, we further employ optimized kinematical cuts according to the kinematical differences
 between the signal and backgrounds to reduce the background to a controlled level.

\vspace{0.7cm}
1. The $2l^\pm l^\mp 2j \rlap/E_T$ signal
\vspace{0.7cm}

Pair production is the main channel of the doubly charged heavy leptons.
Two opposite sign $W$ bosons and two opposite sign leptons are generated by the two heavy leptons $\Sigma^{++}$
and $\Sigma^{--}$ decaying. We demand that one of the $W$ bosons decays leptonically and the other one
decays hadronically. So the final states contain two leptons with same charge, one lepton with opposite charge,
two light jets plus one neutrino,

\begin{eqnarray}
pp \rightarrow \Sigma^{--} \Sigma^{++} \rightarrow l^- W^- l^+ W^+ \rightarrow l^- l^+ j j l^+ \nu (l^- \bar{\nu}),
\end{eqnarray}

The measurement accuracy of the hadronic calorimeter is not enough to distinguish the $W$ or $Z$ boson.
Thereby, the production of the singly charged heavy lepton in association with the doubly charged
heavy lepton also contributes to the above signal. We demand that the $Z$ boson decays to two
light jets with BR$\sim$70$\%$ and the $W$ boson decays leptonically with BR$\sim$21$\%$.
\begin{eqnarray}
pp \rightarrow \Sigma^{\pm\pm} \Sigma^{\mp} \rightarrow l^\pm W^\pm l^\mp Z \rightarrow l^\pm l^\mp j j l^+ \nu(l^- \bar{\nu}).
\end{eqnarray}
Although the production cross section of this channel is smaller than that of the doubly charged heavy lepton pair production
channel, it plays a role in increasing the signal rate. The $2l^\pm l^\mp 2j \rlap/E_T$ signal comes from
both of the processes $pp \rightarrow \Sigma^{--} \Sigma^{++}$ and $pp \rightarrow \Sigma^{\pm\pm} \Sigma^{\mp}$.
The heavy leptons take the same mass as previously described, therefore, we can reconstruct the heavy lepton masses for
these two channels in the same mass range.
The corresponding backgrounds are $l^+l^-2jW^{\pm}$ and $W^+W^-2jW^{\pm}$ where $W$ decays leptonically.

\begin{figure}[htb]
\begin{center}
\includegraphics [scale=0.396] {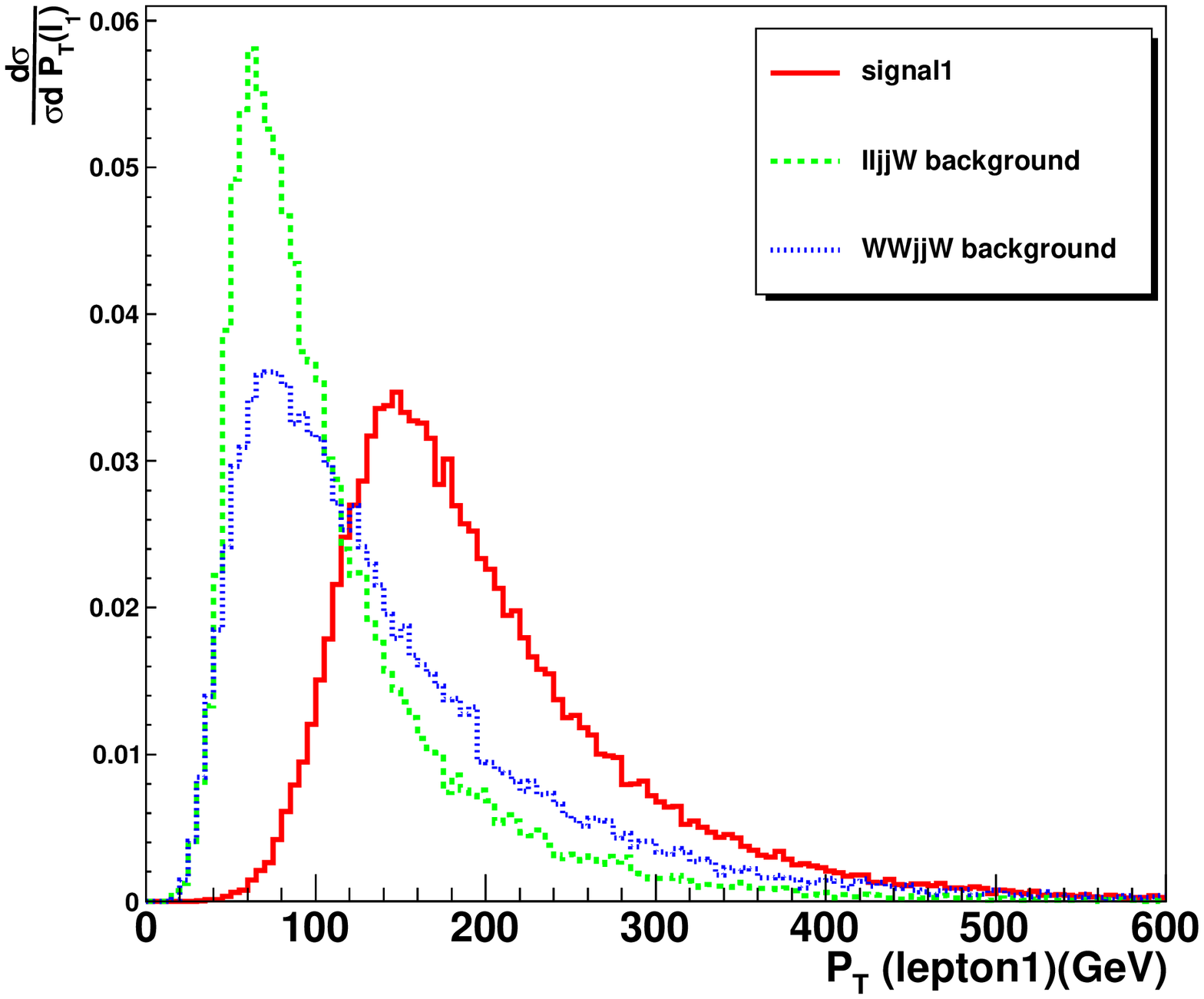}
\includegraphics [scale=0.396] {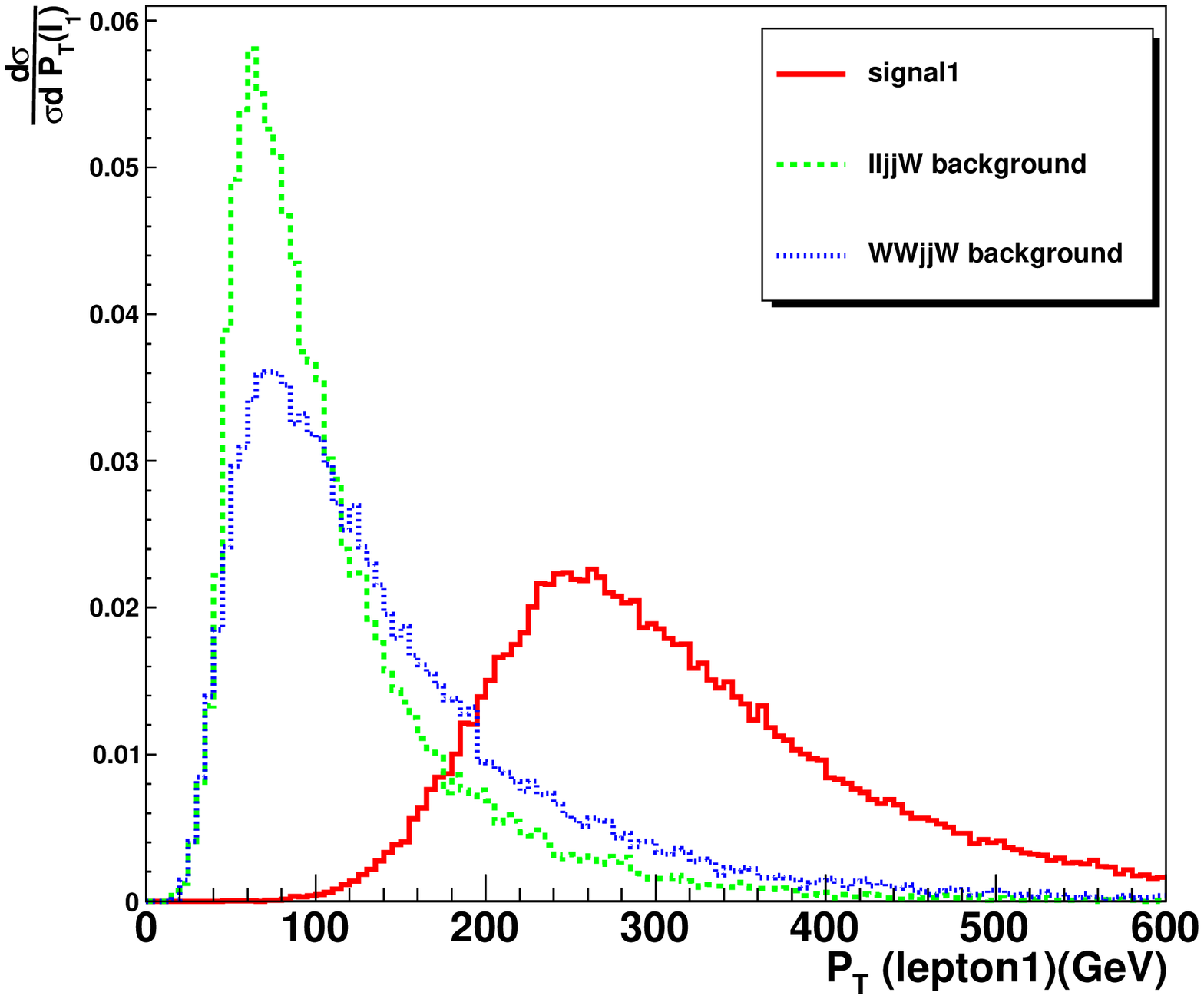}
\vspace{-0.2cm}(a)~~~~~~~~~~~~~~~~~~~~~~~~~~~~~~~~~~~~~~~~~~~~~~~~~~~~~~~~~~~~~~~~~~~~~~~~~~~~~~~~~~(b)
\vspace{-0.2cm}\caption{
Normalized $p_T$ distribution of the leading lepton
in the $2l^\pm l^\mp 2j \rlap/E_T$ signal for \hspace*{1.5cm} $M_\Sigma$=300 GeV (a) and 500 GeV (b) at the 14 TeV LHC.}
\label{Figs:jet1}
\end{center}
\end{figure}

The two jets in the signal events come from $W/Z$ boson decay, however, the jets in the backgrounds mainly come
 from QCD radiation. In order to reduce the background events, we take the invariant mass of the two
jets in the following range,
\begin{eqnarray}
M_W-20 {\rm GeV} < M(jj)<M_Z+20 {\rm GeV}.
\end{eqnarray}

\begin{figure}[htbp]
\begin{center}
\includegraphics [scale=0.396] {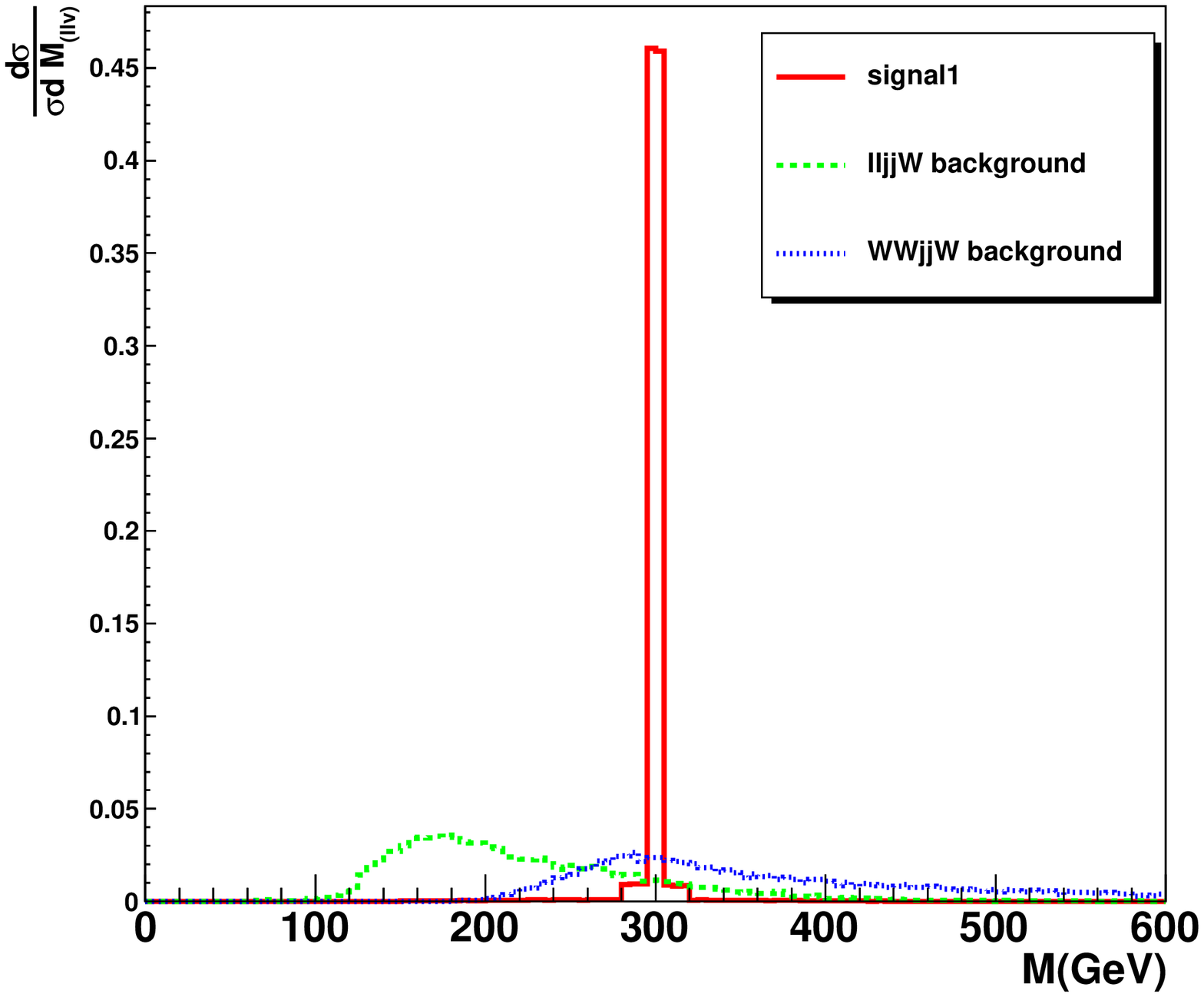}
\includegraphics [scale=0.396] {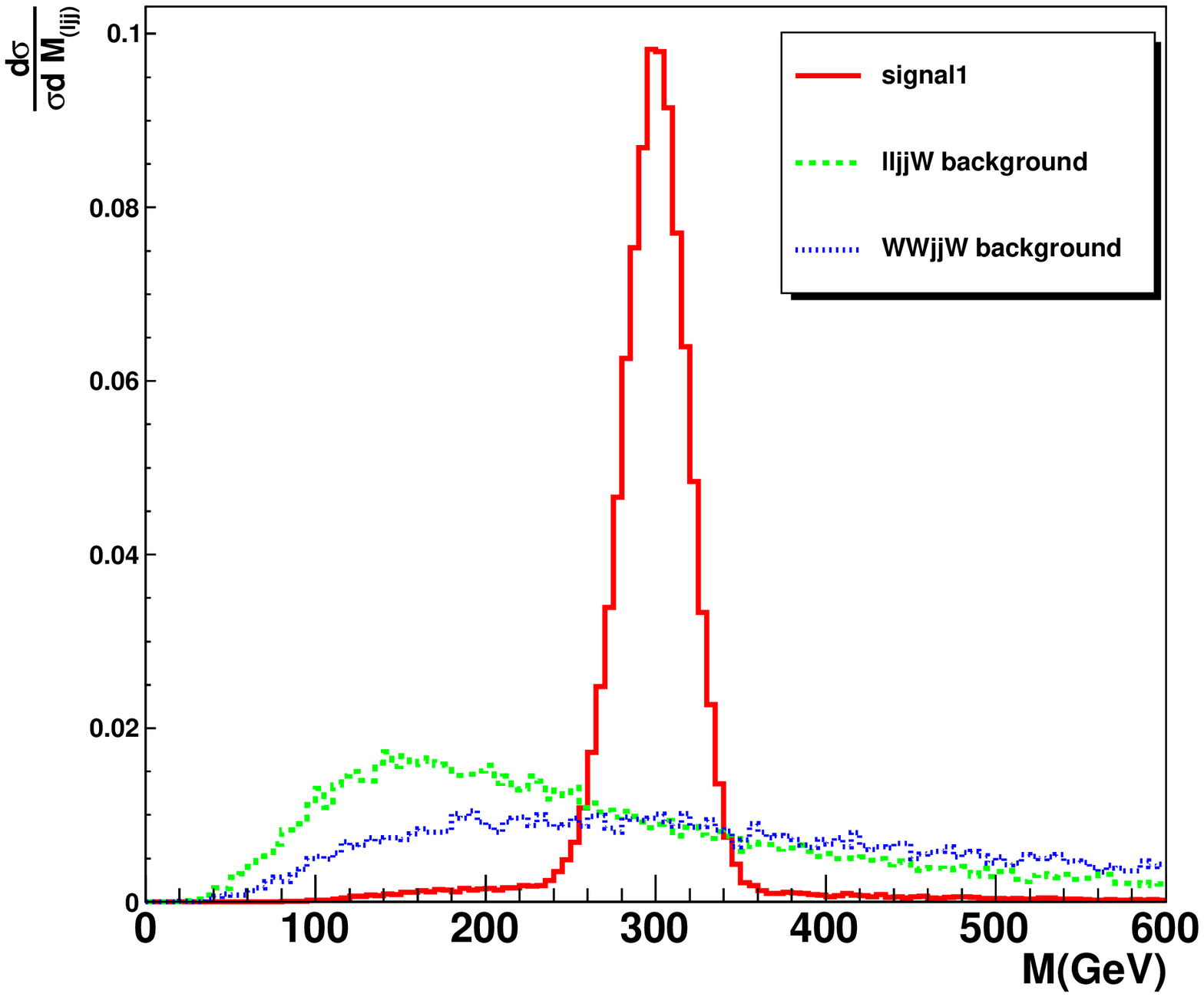}
\vspace{-0.2cm}(a)~~~~~~~~~~~~~~~~~~~~~~~~~~~~~~~~~~~~~~~~~~~~~~~~~~~~~~~~~~~~~~~~~~~~~~~~~~~~~~~~~~(b)\\
\vspace{0.5cm}
\includegraphics [scale=0.396] {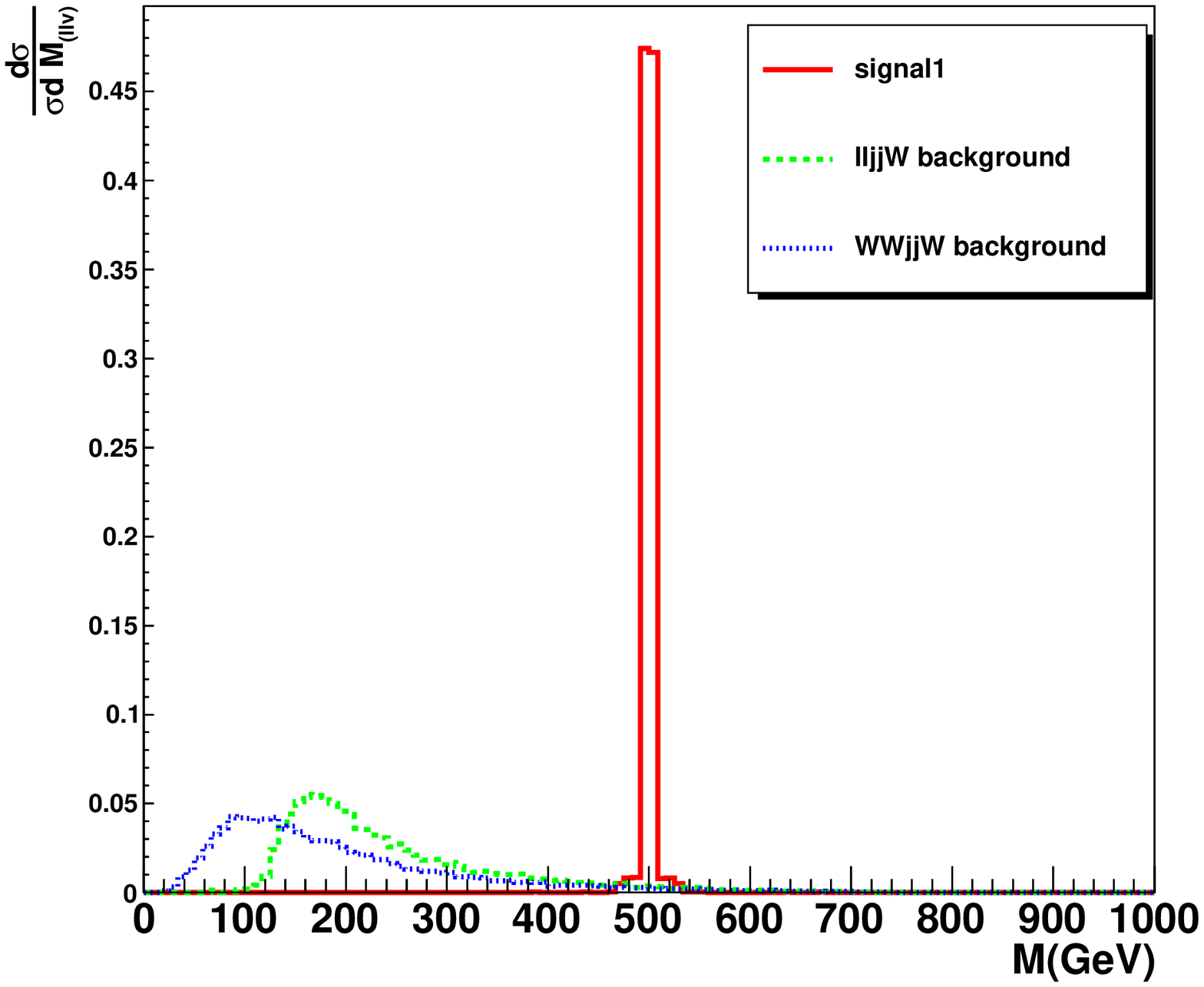}
\includegraphics [scale=0.396] {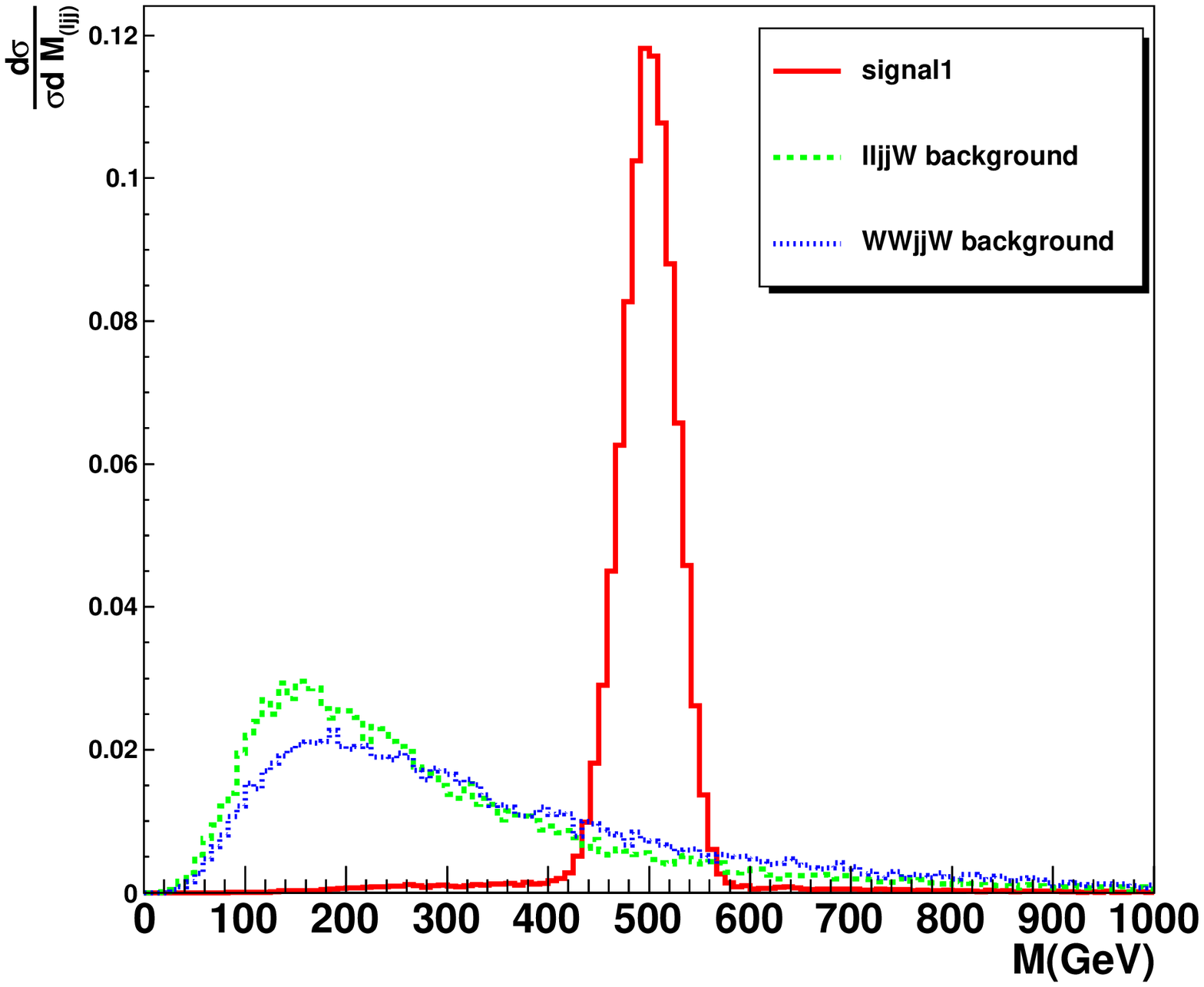}
\vspace{-0.2cm}(c)~~~~~~~~~~~~~~~~~~~~~~~~~~~~~~~~~~~~~~~~~~~~~~~~~~~~~~~~~~~~~~~~~~~~~~~~~~~~~~~~~~(d)
\vspace{-0.2cm}\caption{
Normalized invariant mass distribution of $M(llv)$ and $M(ljj)$ in the $2l^\pm l^\mp 2j \rlap/E_T$ signal
 \hspace*{1.3cm}for $M_\Sigma$=300 GeV (a,b) and 500 GeV (c,d) at the 14 TeV LHC.}
\label{Figs:jet1}
\end{center}
\end{figure}

As discussing above, there are three SM leptons in the final states.
The lepton which has the largest transverse momentum is defined as
the leading lepton (lepton1), it comes from the heavy lepton decay in the signal. Its $p_T$ spectrum peaks at
around half of the heavy lepton mass while the lepton in the background tends to be soft.
We order the leptons by their values of $p_T$ for the signal and backgrounds, and display the normalized $p_T$
distribution of the leading lepton (lepton1) for the $2l^\pm l^\mp 2j \rlap/E_T$ signal and background events in Fig.2.
For the leading lepton (lepton1), we can see that the signal distribution (the red solid line) peaks at around 150 (250) GeV
for the heavy lepton mass $M_\Sigma$=300 (500) GeV while the $lljjW$ (green dashed line) and $WWjjW$ (the blue dotted line)
 background distributions peak at around 80 GeV. We can distinguish between the signal and the backgrounds
by a cut based on the kinematical variable $p_T$ of the leading lepton (lepton1) as follows,
\begin{eqnarray}
p_T(lepton1) > 100 (160){\rm GeV},
\end{eqnarray}
where the cut $p_T$(lepton1) $> 100$ GeV corresponds to $M_\Sigma$=300 GeV and the value in parenthesis is
the case for $M_\Sigma$=500 GeV. The cuts are very effective in reducing the backgrounds and preserving
the signal events.

We subsequently reconstruct the masses of the heavy leptons to further suppress the backgrounds.
The two jets with one charged lepton in the final states can reconstruct one heavy lepton
mass $M(ljj)$, and the remaining two charged leptons and one neutrino can reconstruct
another heavy lepton mass $M(ll\nu)$.
The normalized invariant mass distribution of the two heavy leptons after the basic cuts are shown in Fig.3.
We can see that the SM background events distribute in the low invariant mass region.
The invariant masses of the heavy leptons
in the signal events are larger than those in the background events for both $M_\Sigma$=300 GeV and 500 GeV.
Because we are ignoring the mass splitting among the heavy leptons, we choose the solution which gives the
closest $M(ll\nu)$ to $M(ljj)$ and take the detailed cut as follows,
\begin{eqnarray}
|M(ll\nu)-M(ljj)|<30(50){\rm GeV}.
\end{eqnarray}

\begin{table}
\begin{center}
\begin{tabular}{|c|c|c|c|}\hline
   &\multicolumn{1}{|c|}{Signal $2l^\pm l^\mp 2j \rlap/E_T$}
   &\multicolumn{1}{|c|}{Bkg $l^+l^-2jW^\pm$}
   &\multicolumn{1}{|c|}{Bkg $W^+W^-2jW^\pm$}  \\ \hline
Basic cuts                             &$28.92$ $(4.14$)       &$56.91$             &$0.308$  \\ \hline
60GeV$<M_{jj}<$110GeV                   &$27.81$ $(4.03)$       &$12.72$             &$0.058$     \\ \hline
$p_T(l_1)>$100(160)GeV                &$26.41$ $(3.89)$         &$ 4.63$ $(1.62)$     &$0.031$ $(0.015)$     \\ \hline
$|M_{ll\nu}-M_{ljj}|<30$(50)GeV           &$22.44$ $(3.56)$       &$ 0.42$ $(0.36) $   &$0.005$ $(0.003)$    \\ \hline
Number of events               &$224.4$ $(35.6)$               &$4.2$ $(3.6)$       &$0.05$ $(0.03)$  \\ \hline
$S/\sqrt{S+B}$                 &\multicolumn{3}{|c|}{14.84 (5.68)}               \\ \hline
\end{tabular}
\caption{\small  The cross sections (fb) and the event numbers of the signal $2l^\pm l^\mp jj \rlap/E_T$ and the back
                 \hspace*{1.9cm}grounds $l^+l^-2jW^\pm$ and $W^+W^-2jW^\pm$ for $M_\Sigma$=300 (500) GeV at the 14 TeV LHC
                  \hspace*{1.9cm}with $\mathcal{L}=10$ fb$^{-1}$.}
\end{center}
\end{table}

\begin{figure}[htb]
\begin{center}
 \epsfig{file=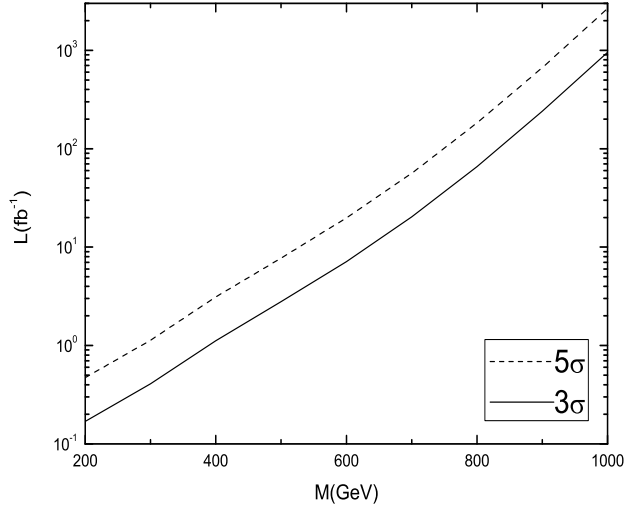,width=270pt,height=240pt}
\caption{The needed luminosity to observe different masses of the heavy leptons via the $2l^\pm l^\mp 2j \rlap/E_T$
\hspace*{1.3cm} signal for 3$\sigma$ and $5\sigma$ statistical significances at the 14 TeV LHC.} \label{feynman:ww}
\end{center}
\end{figure}

After all of these cuts are applied, the cross sections of this signal and the backgrounds are listed in Table I.
The former data in every columns and the data in the parentheses correspond to the results for the
heavy lepton mass $M_\Sigma=300$ GeV and $M_\Sigma=$500 GeV, respectively.
It is obvious that the sets of cuts can significantly suppress the backgrounds.
We define the statistical significance  as $s=S/{\sqrt{(S+B)}}$ where $S$ and $B$ denote the number
of signal and background events, respectively.
It can reach 14.84 (5.68) for $M_\Sigma$=300 (500) GeV at the 14 TeV LHC with an integrated luminosity of 10 fb$^{-1}$.
In order to illustrate the needed integrated luminosity at LHC to reach a given statistical significance,
we plot the integrated luminosity versus the heavy lepton mass for $3\sigma$ and $5\sigma$ statistical significances for the $2l^\pm l^\mp 2j \rlap/E_T$ signal at the 14 TeV LHC in Fig.4.
As is shown in Fig.4., this signal can be detected at the 14 TeV LHC under the designed integrated luminosity in most mass ranges of the heavy lepton. For $M_\Sigma$=500 (700) GeV, the 5$\sigma$ significance requires 7.74 (56.32) fb$^{-1}$.


\vspace{0.7cm}
2. The $2l^\pm 2 l^\mp 2j $ signal
\vspace{0.7cm}

The production of the singly charged heavy leptons in association with the doubly charged heavy leptons
$\Sigma^{\pm\pm}\Sigma^\mp$ can provide a distrinct signal in a case where $W$ decays hadronically and $Z \rightarrow l^+l^-$.
Thus, there are two positively charged leptons, two negatively charged leptons plus two jets in the final states,

\begin{eqnarray}
pp \rightarrow \Sigma^{\pm\pm} \Sigma^{\mp} \rightarrow l^\pm W^\pm l^\mp Z \rightarrow l^\pm l^\pm l^\mp l^\mp j j .
\end{eqnarray}

\begin{figure}[htbp]
\begin{center}
\includegraphics [scale=0.396] {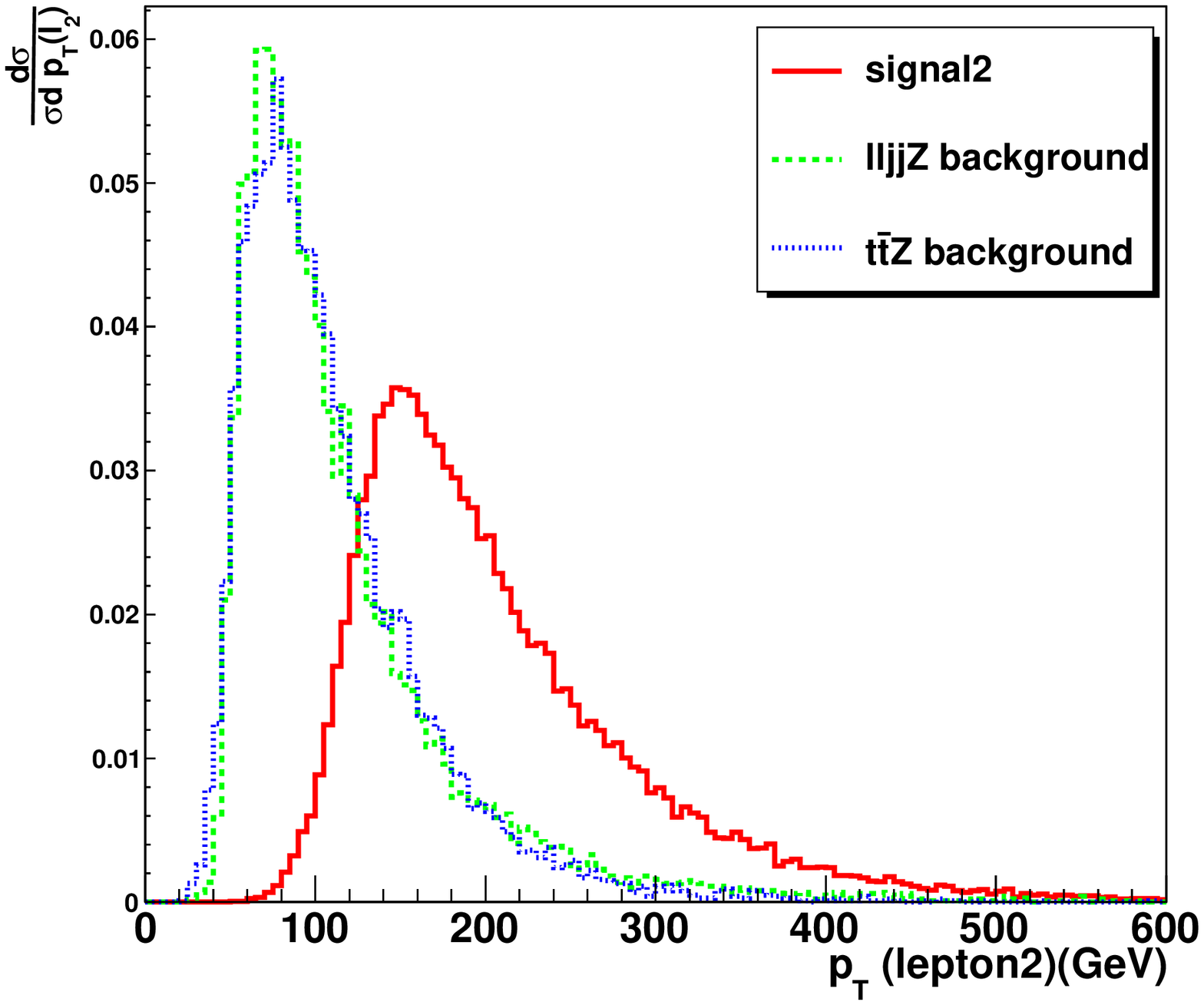}
\includegraphics [scale=0.396] {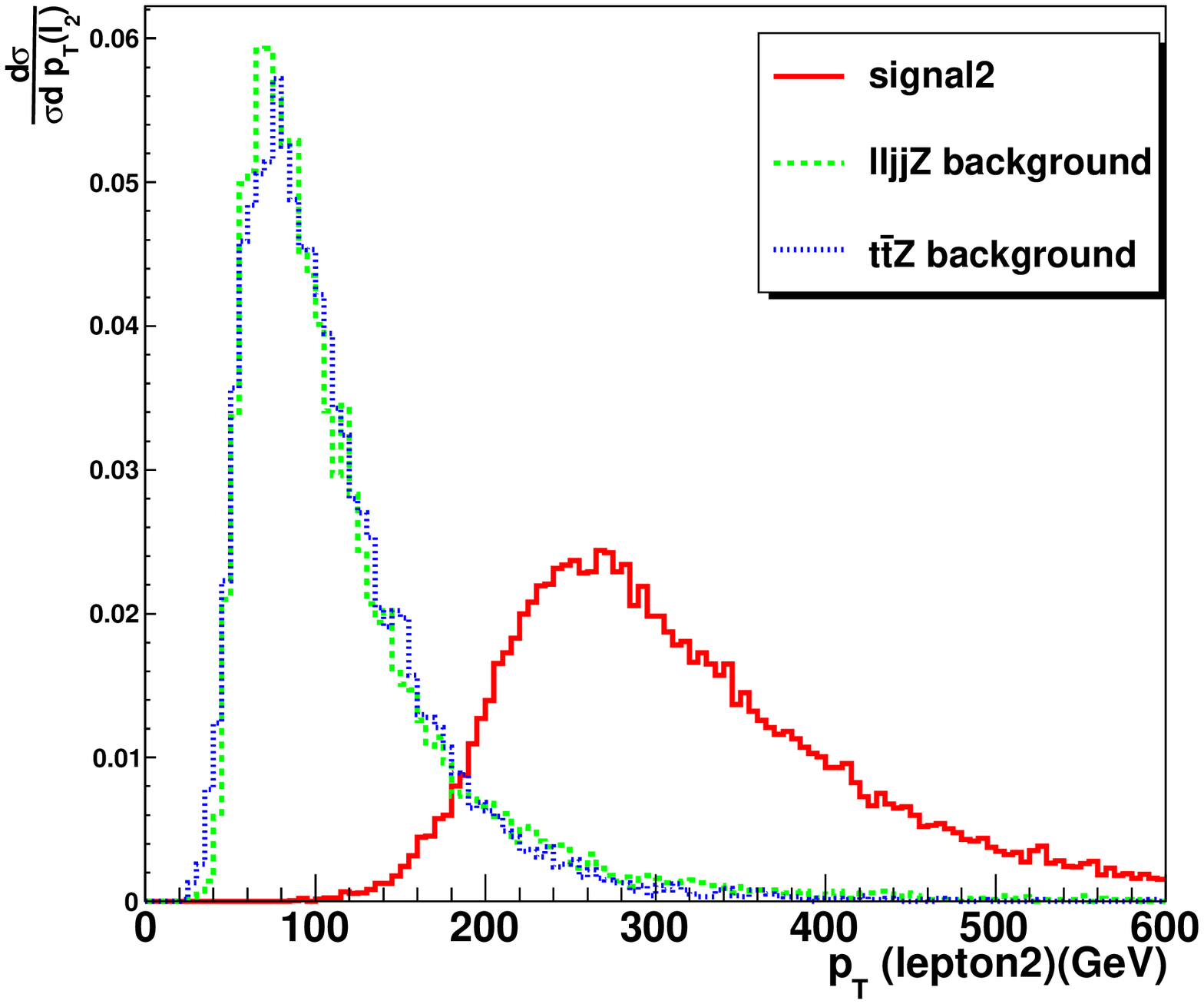}
\vspace{-0.2cm}(a)~~~~~~~~~~~~~~~~~~~~~~~~~~~~~~~~~~~~~~~~~~~~~~~~~~~~~~~~~~~~~~~~~~~~~~~~~~~~~~~~~~(b)
\vspace{-0.2cm}\caption{
Normalized $p_T$ distribution of the leading lepton
in the $2l^\pm 2 l^\mp 2j $ signal for $M_\Sigma=$ \hspace*{1.3cm}300 GeV (a) and 500 GeV (b) at the 14 TeV LHC. }
\label{Figs:jet1}
\end{center}
\end{figure}

\begin{figure}[!htb]
\begin{center}
\includegraphics [scale=0.396] {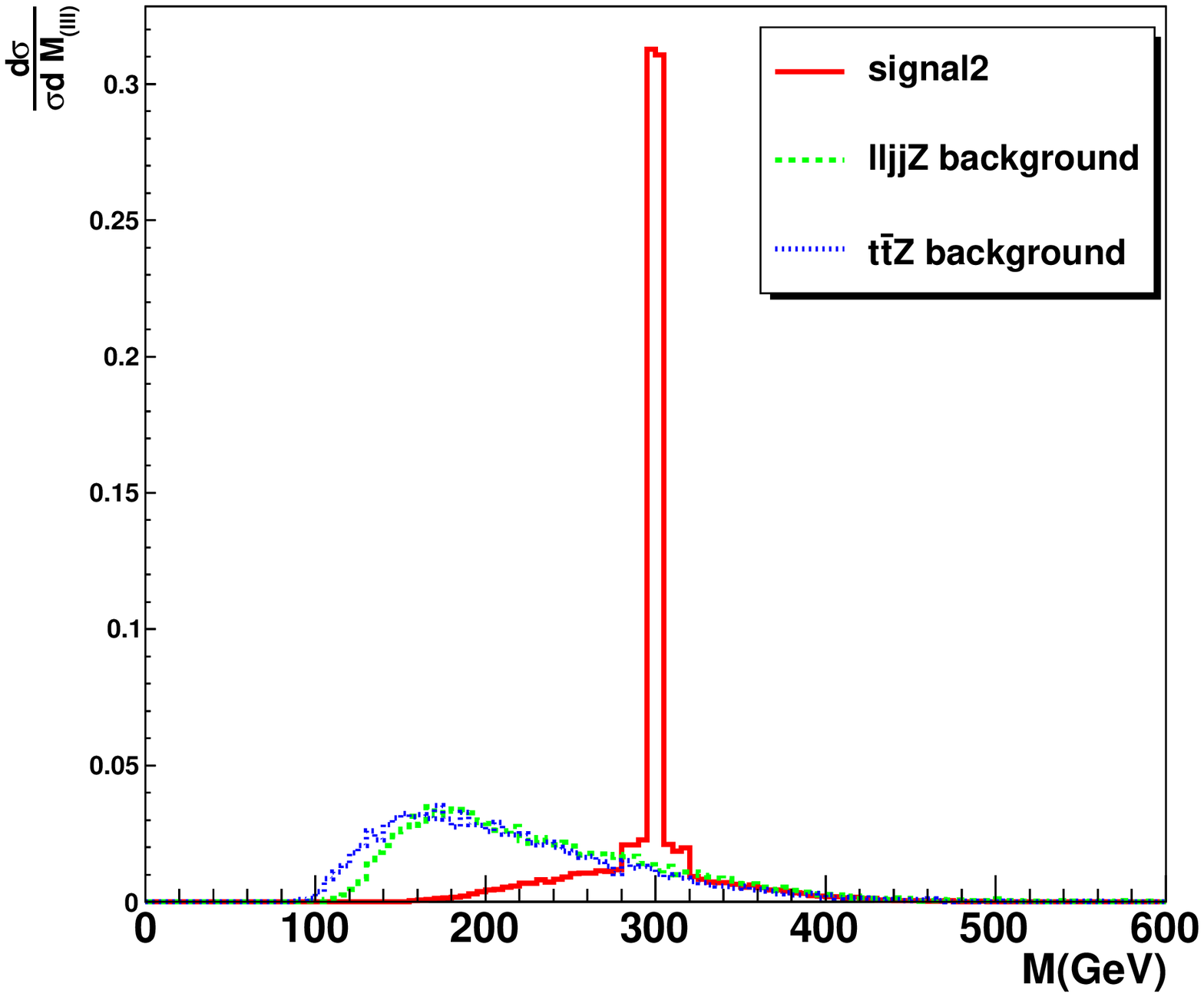}
\includegraphics [scale=0.396] {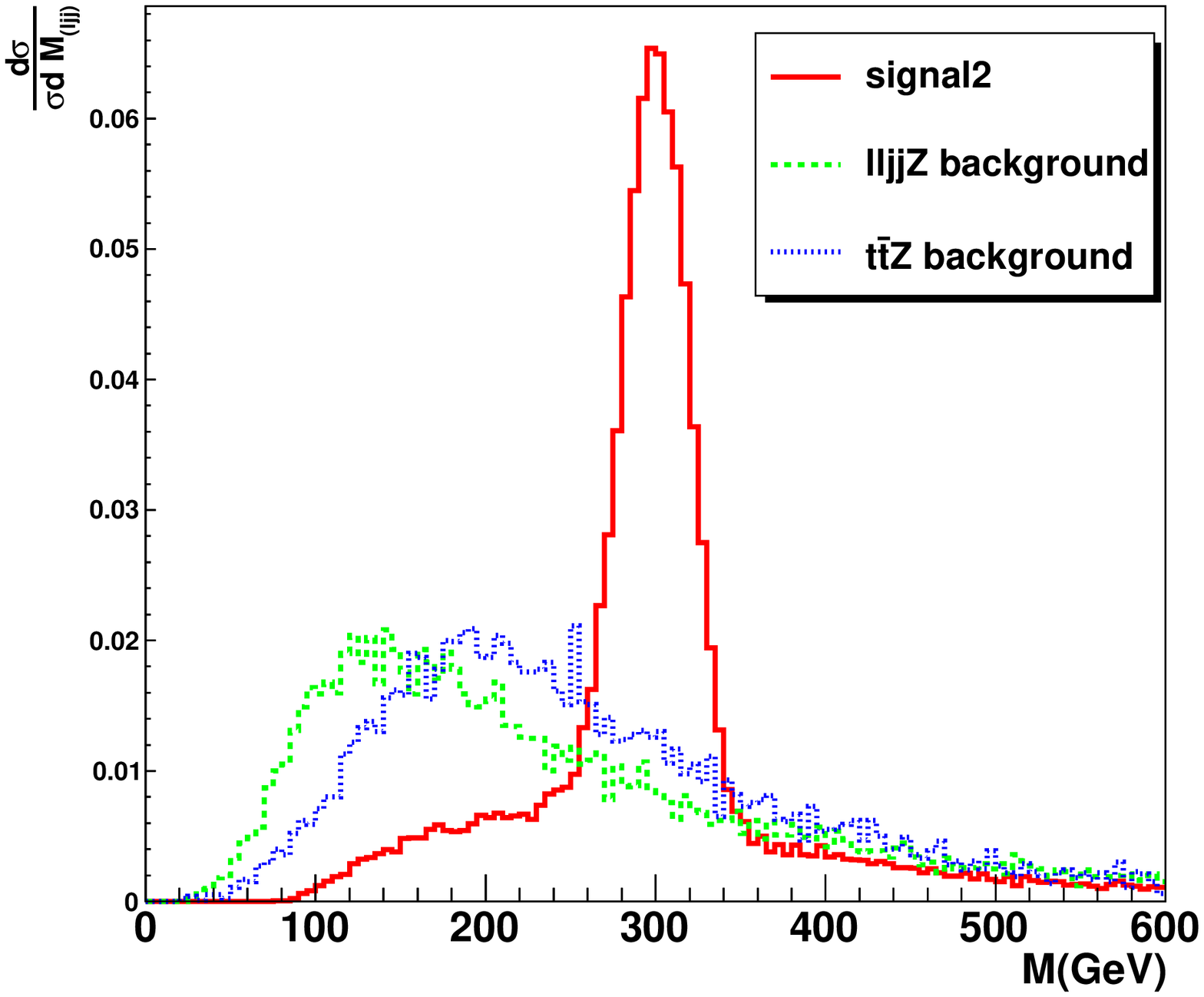}
\vspace{-0.2cm}(a)~~~~~~~~~~~~~~~~~~~~~~~~~~~~~~~~~~~~~~~~~~~~~~~~~~~~~~~~~~~~~~~~~~~~~~~~~~~~~~~~~~(b)\\
\vspace{0.5cm}
\includegraphics [scale=0.396] {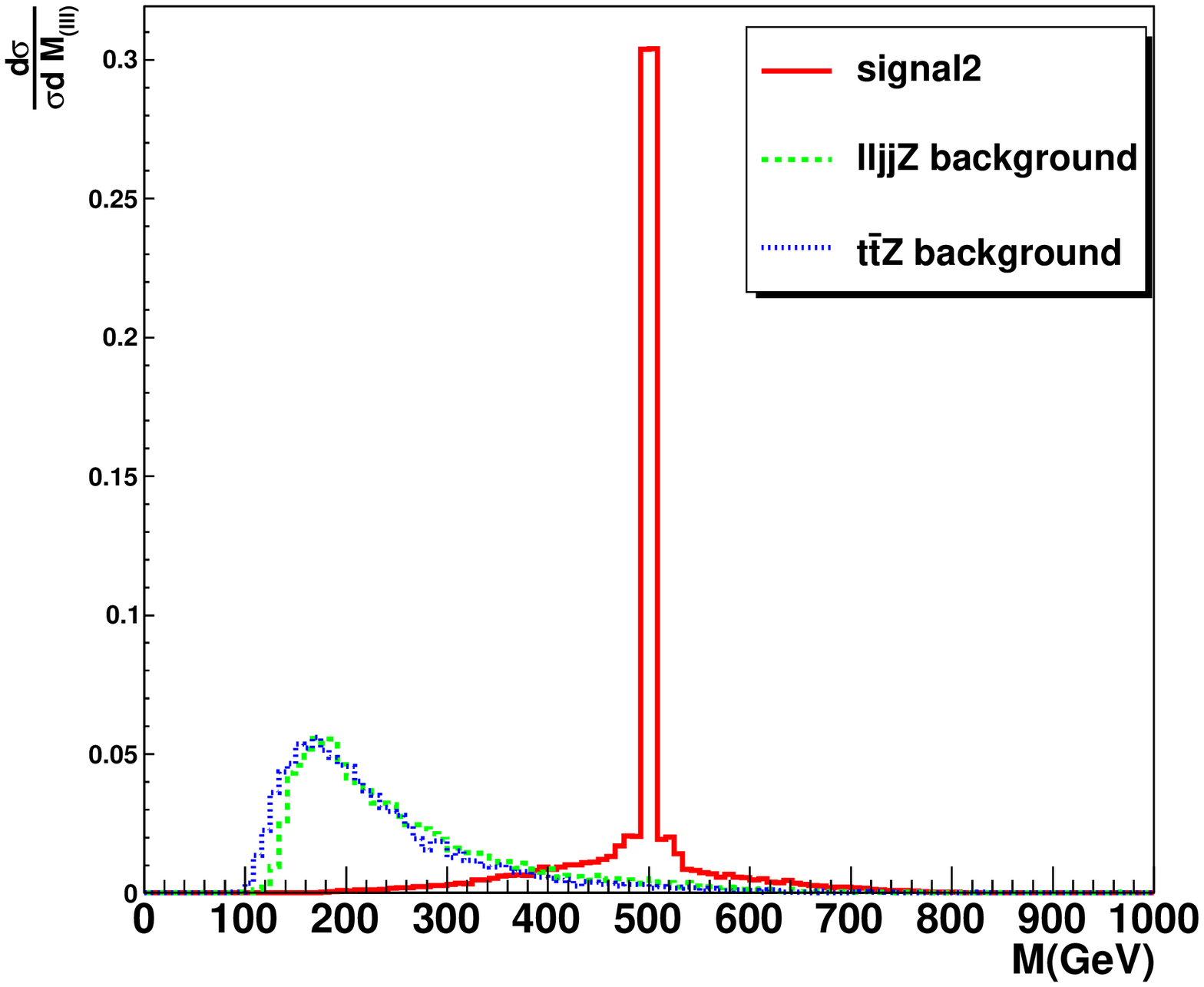}
\includegraphics [scale=0.396] {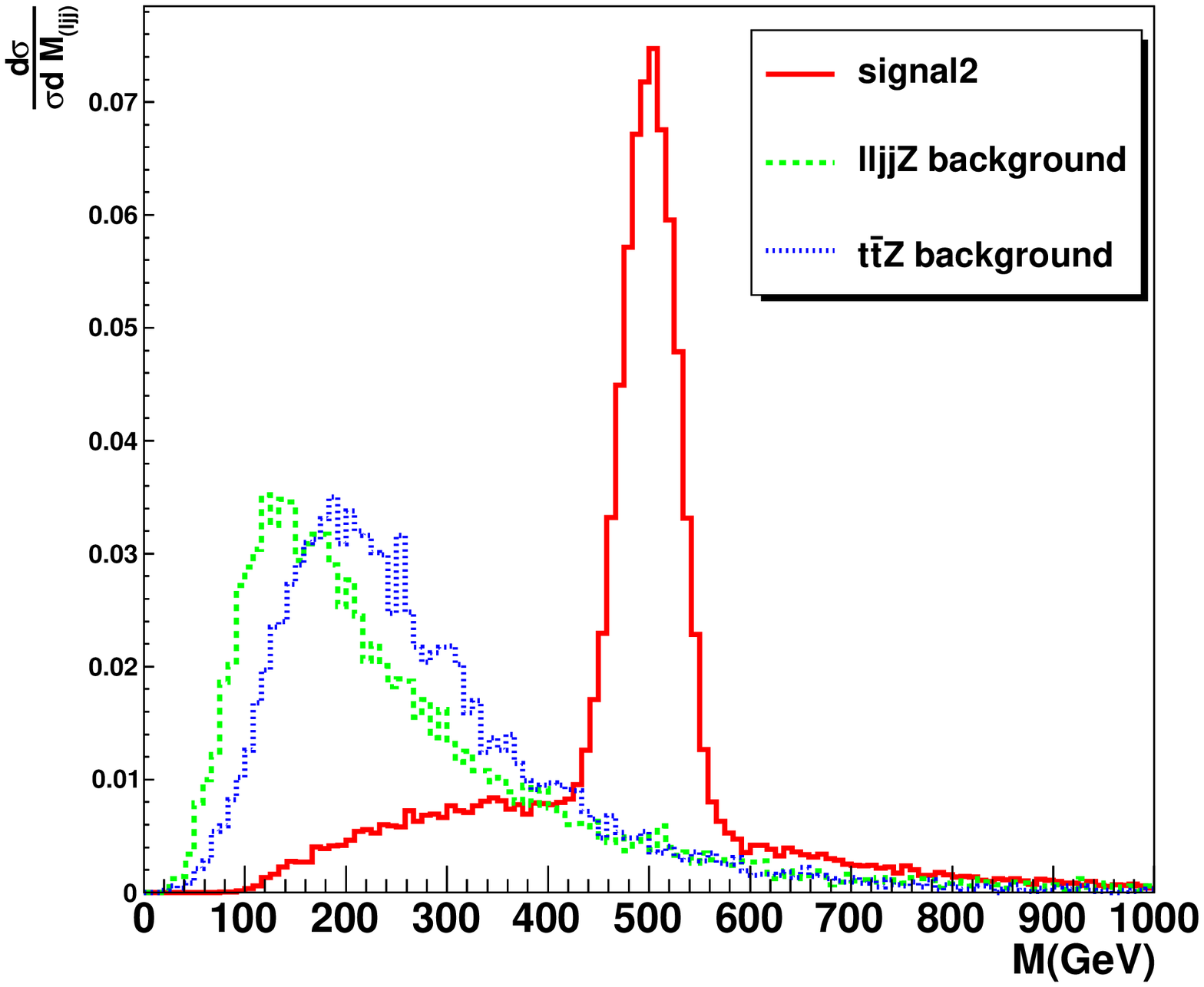}
\vspace{-0.2cm}(c)~~~~~~~~~~~~~~~~~~~~~~~~~~~~~~~~~~~~~~~~~~~~~~~~~~~~~~~~~~~~~~~~~~~~~~~~~~~~~~~~~~(d)
\vspace{-0.2cm}\caption{
Normalized invariant mass distribution of $M(lll)$ and $M(ljj)$ in the $2l^\pm l^\mp 2j \rlap/E_T$ signal
 \hspace*{1.3cm}for $M_\Sigma$=300 GeV (a,b) and 500 GeV (c,d) at the 14 TeV LHC.}
\label{Figs:jet1}
\end{center}
\end{figure}

Another two channels also contribute to the signal. For the pair production channel
 $\Sigma^{\pm} \Sigma^{\mp}$, one of the $Z$ bosons
 decays hadronically and another decays to $l^+ l^-$. The decay modes of $W$ and $Z$ in $\Sigma^{\pm} \Sigma^0 $
 production are consistent with $\Sigma^{\pm\pm} \Sigma^{\mp}$ production mentioned above,
\begin{eqnarray}
pp \rightarrow \Sigma^{\pm} \Sigma^{\mp} \rightarrow l^\pm Z l^\mp Z \rightarrow l^\pm l^\pm l^\mp l^\mp j j ,
\end{eqnarray}
\begin{eqnarray}
pp \rightarrow \Sigma^{\pm} \Sigma^0 \rightarrow l^\pm Z l^\mp W^\pm \rightarrow l^\pm l^\pm l^\mp l^\mp j j.
\end{eqnarray}


The corresponding backgrounds are $l^+l^-2jZ$ and $t\bar{t}Z$, where $Z \rightarrow l^+l^-$ and $t \rightarrow bl^+\nu$
($\bar{t}\rightarrow \bar{b} l^- \bar{\nu}$).
  As there is no neutrino in the signal
but there are neutrinos in the backgrounds, we apply a veto cut about the missing transverse
momentum $\rlap/E_T<25$ GeV replacing that in the basic cuts to reduce the $t\bar{t} Z$ events.
We also require the invariant mass of the two jets to peak at the $W$/$Z$ mass within a mass
 window of 20 GeV. This cut can rapidly reduce the background
 while affecting the signal slightly,

\begin{eqnarray}
M_W-20{\rm GeV} < M(jj)<M_Z+20{\rm GeV}.
\end{eqnarray}

We also plot the normalized $p_T$ distribution of the leading lepton (lepton2) for
the $2l^\pm 2l^\mp 2j$ signal and background events for $M_\Sigma$=300 GeV and 500 GeV in Fig.5.
The same cuts based on the transverse momentum
$p_T$ as mentioned in the $2l^\pm l^\mp 2j \rlap/E_T$ signal are applied to suppress the
backgrounds and strengthen the signal,
\begin{eqnarray}
p_T(lepton2) > 100(160){\rm GeV}.
\end{eqnarray}

\begin{table}
\begin{center}
\begin{tabular}{|c|c|c|c|}\hline
   &\multicolumn{1}{|c|}{Signal $2l^\pm 2 l^\mp 2j $}
   &\multicolumn{1}{|c|}{Bkg $l^+l^-2jZ$}
   &\multicolumn{1}{|c|}{Bkg $t\bar{t}Z$}  \\ \hline
Basic cuts                                &$0.530$ $(7.68\times 10^{-2})$     &$ 5.661 $             &$ 0.071$             \\ \hline
60GeV$<M_{jj}<$110GeV                     &$0.511$ $(7.49\times 10^{-2})$     &$1.429$       &$0.013$     \\ \hline
$p_T(l_2)>$ 100(160)GeV                   &$0.501$ $(7.39\times 10^{-2})$     &$0.553$   (0.168)     &$0.006$ (0.002)     \\ \hline
$|M_{ll\nu}-M_{ljj}|<30$(50)GeV           &$0.328$ $(5.18\times 10^{-2})$     &$0.074$  (0.024)     &$0.001$ (0)   \\ \hline
Number of events              &$32.8$ (5.18)                      &$7.4 $ (2.4)            &$0.1$ (0) \\ \hline
$S/\sqrt{S+B}$                 &\multicolumn{3}{|c|}{5.17 (1.87)}           \\ \hline
\end{tabular}
\caption{\small   The cross sections (fb) and the event numbers of the signal $2l^\pm 2l^\mp 2j $ and the
\hspace*{2cm} backgrounds $l^+l^-2jZ$ and $t\bar{t}Z$ for $M_\Sigma$=300 (500) GeV at the 14 TeV LHC with
\hspace*{2.2cm} $\mathcal{L}$=100 fb$^{-1}$.}
\end{center}
\end{table}

\begin{figure}[htb]
\begin{center}
 \epsfig{file=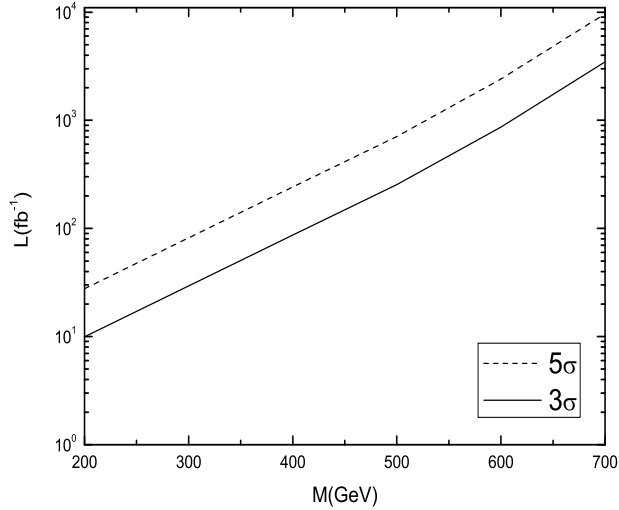,width=270pt,height=240pt}
\caption{The needed luminosity to observe different mass quintuplet leptons via the $2l^\pm 2 l^\mp 2j $
\hspace*{1.3cm}signal for the 3$\sigma$ and $5\sigma$ statistical significances at the 14 TeV LHC.} \label{feynman:ww}
\end{center}
\end{figure}

One can reconstruct the mass of one heavy lepton via three light leptons, and the other one via the remmant
lepton and the two jets. We plot the normalized invariant mass of the
two heavy leptons $M_{lll}$ and $M_{ljj}$ in Fig.6. In order to further suppress the background to
manageable levels, the same invariant mass cuts as were used for the $2l^\pm l^\mp 2j \rlap/E_T$ signal are applied
to the signal and the backgrounds,
\begin{eqnarray}
|M(lll)-M(ljj)|<30(50){\rm GeV}.
\end{eqnarray}

We summarize the results in Table II. The cross section of $t\bar{t}Z$
is too small for $M_\Sigma=500$ GeV, we consider that it is approximately zero.
The cross sections of the backgrounds are tiny compared to the signal after sequential cuts,
and the statistical significance $s$ can reach 5.17 (1.87) at the 14 TeV LHC with
 an integrated luminosity of 100 fb$^{-1}$. We also give the integrated luminosity versus the
 heavy lepton mass for $3\sigma$ and $5\sigma$ statistical significance for the $2l^\pm 2 l^\mp 2j $ signal at the
14 TeV LHC in Fig.7. If we want to observe this signal for a $5\sigma$ statistical significances at
$M_\Sigma=300$ (500) GeV, the integrated luminosities must be larger than 81.797 (706.236) fb$^{-1}$ at 14 TeV LHC.
For $M_\Sigma>700$ GeV, detecting this signal at 5$\sigma$ requires an integrated luminosity larger than $10^4$, which outreaches the designed luminosity.


\vspace{0.7cm}

3. The $3l^\pm l^\mp 2j$, $3l^\pm 2l^\mp \rlap/E_T$ and $3l^\pm 3l^\mp$ signals

\vspace{0.7cm}

The lepton-number violating (LNV) processes have a clean SM background and they are easily detected in the experiments \cite{sourceLNV}.
 In this paper, LNV like-sign dilepton events are mediated by the exotic heavy lepton decaya and are
reminiscent of those found in related canonical seesaw models like the type III seesaw \cite{LNV}.
They all predict the $l^\pm l^\pm W^\mp Z$ events. We can get the $3l^\pm l^\mp 2j$ signal after the $W$ and $Z$ boson decays,
which resemble the $2l^\pm 2 l^\mp 2j $ signal.
\begin{eqnarray}
pp \rightarrow \Sigma^{\pm} \Sigma^0 \rightarrow l^\pm Z l^\pm W^\mp \rightarrow l^\pm l^\pm l^\pm l^\mp j j .
\end{eqnarray}
It is obvious that the leptons with opposite charge can be distinguished in the experiments. Thus, this channel provides a different signal
for observing the heavy leptons compared to $2l^\pm 2 l^\mp 2j $ signal. $W^\pm W^\pm Z jj$ is treated as the background in which $W$ decays leptonically
and $Z \rightarrow l^+ l^-$. The cross section of the background is much smaller than that of the signal, we only
apply the basic cuts on the signal and the background. All of the results are listed in Table III.

We also consider the $3l^\pm 2l^\mp \rlap/E_T$ signal which is generated by $\Sigma^{\pm\pm}\Sigma^\mp$
 and $\Sigma^\pm \Sigma^0$ productions,

\begin{eqnarray}
pp \rightarrow \Sigma^{\pm\pm} \Sigma^{\mp} \rightarrow l^\pm W^\pm  l^\mp Z \rightarrow l^\pm l^\pm l^\pm l^\mp l^\mp \nu (\bar{\nu}),
\end{eqnarray}
\begin{eqnarray}
pp \rightarrow \Sigma^{\pm} \Sigma^0 \rightarrow l^\pm Z l^\mp W^\pm (l^\pm Z l^\pm W^\mp) \rightarrow l^\pm l^\pm l^\pm l^\mp l^\mp \bar{\nu}(\nu) ,
\end{eqnarray}
\begin{eqnarray}
pp \rightarrow \Sigma^{\pm} \Sigma^0 \rightarrow l^\pm Z Z \nu \rightarrow l^\pm l^\pm l^\pm l^\mp l^\mp \nu .
\end{eqnarray}
where the $l^\pm Z l^\pm W^\mp$ and $l^- Z Z \nu$ events are all from the LNV heavy lepton decays and the subsequent leptonic
$Z/W$ decay. The relevant background is $ZZW^\pm$.
The production cross section of the signal is large enough compared to the small background, which is similar to the
$3l^\pm l^\mp 2j$ signal.
All of the results that apply to the basic cuts are displayed in Table III. We also calculate the
the needed integrated luminosity to observe different mass quintuplet leptons via the $3l^\pm l^\mp 2j$ and
$3l^\pm 2l^\mp \rlap/E_T$ signals at the 14TeV LHC in Fig.8.
For $M_\Sigma$=300 (500) GeV, the 5$\sigma$ significance requires 139.2 (1009.1) fb$^{-1}$.

\begin{table}[!htb]
\begin{center}
\begin{tabular}{|c|c|c|c|}\hline
   &\multicolumn{1}{|c|}{Signal $3l^\pm l^\mp 2j$}
   &\multicolumn{1}{|c|}{Bkg $W^\pm W^\pm Z 2j$}
    &\multicolumn{1}{|c|}{$S/\sqrt{S+B}$ }\\ \hline
Basic cuts                    &$0.208$ ($3.01\times 10^{-2}$)     &$ 1.09 \times 10^{-3}$     &$-$ \\ \hline
Number of events       &$ 20.8$ (3.01)                     &$ 0.11 $                   &$4.54$ (1.70)      \\\hline\hline
   &\multicolumn{1}{|c|}{Signal $3l^\pm 2l^\mp \rlap/E_T$}
   &\multicolumn{1}{|c|}{Bkg $ZZW^\pm$}
    &\multicolumn{1}{|c|}{$S/\sqrt{S+B}$ }\\ \hline
Basic cuts                    &$0.184$ $(2.41\times 10^{-2})$           &$4.51 \times 10^{-3}$  &$-$\\ \hline
Number of events       &$ 18.4 $ ($2.41$)                         &$ 0.45 $               &$4.24$ (1.43)    \\ \hline
\end{tabular}
\caption{\small  The cross sections (fb) and the event numbers of the signals $3l^\pm l^\mp 2j$, $3l^\pm 2l^\mp \rlap/E_T$ \hspace*{2cm}
and the SM backgrounds for $M_\Sigma=300$ (500) GeV at the 14 TeV LHC with
 \hspace*{2cm} $\mathcal{L}$=100 fb$^{-1}$.}
\end{center}
\end{table}

\begin{figure}[htb]
\begin{center}
\includegraphics [scale=0.7] {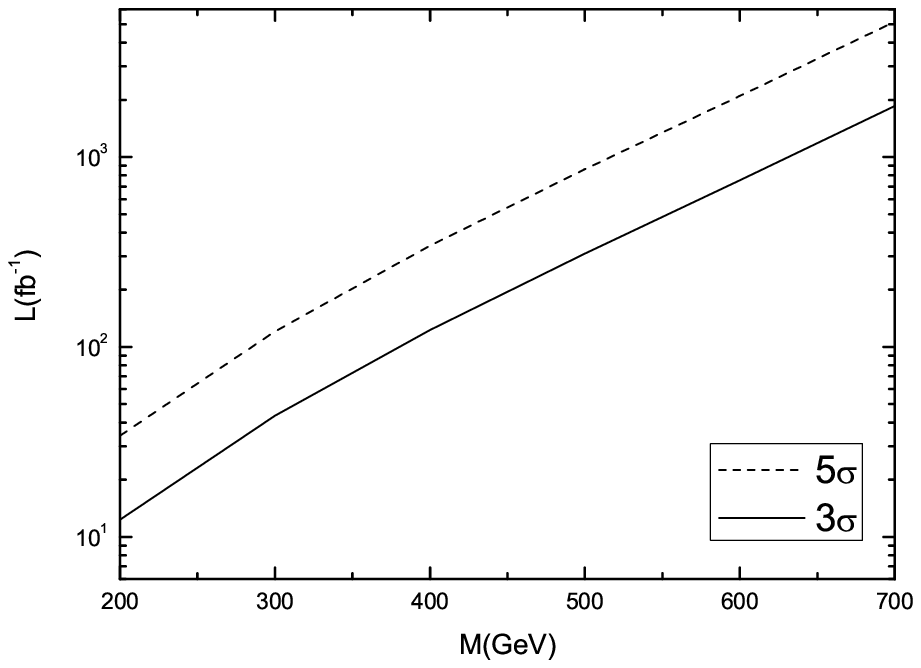}
\includegraphics [scale=0.7] {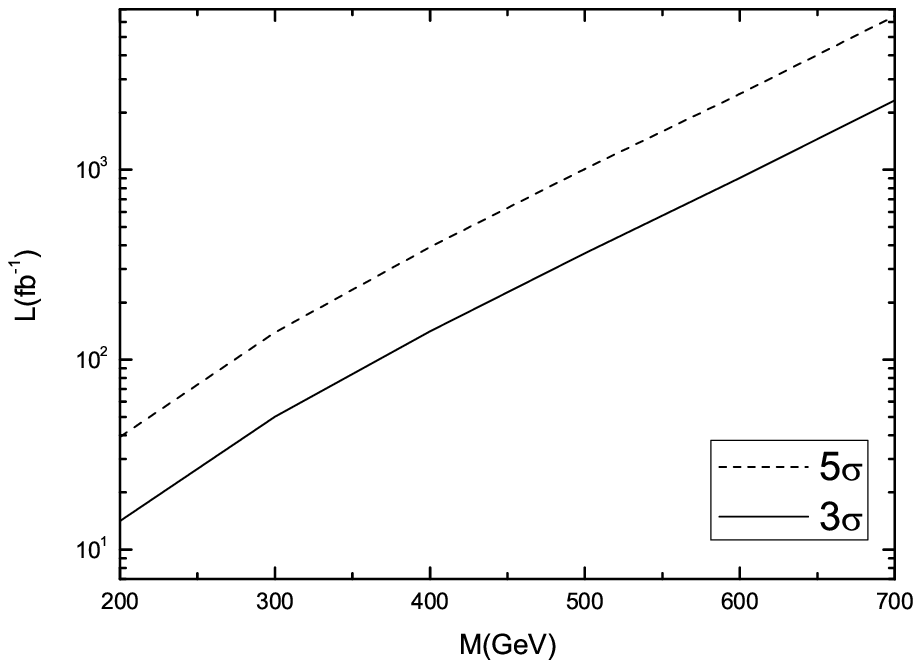}
\vspace{-0.2cm}
\vspace{-0.2cm}\caption{
The needed luminosity to observe different mass quintuplet leptons via $3l^\pm l^\mp 2j $(left \hspace*{1.3cm} panel) and $3l^\pm 2l^\mp \rlap/E_T$(right panel) signals  for 3$\sigma$ and $5\sigma$ statistical significances at the \hspace*{1.3cm} 14 TeV LHC.}
\label{Figs:jet1}
\end{center}
\end{figure}

The last signal considered is a clean channel which consists of six leptons in the final states,
\begin{eqnarray}
pp \rightarrow \Sigma^\pm \Sigma^\mp \rightarrow l^\pm Z l^\mp Z \rightarrow l^\pm l^\pm l^\pm l^\mp l^\mp l^\mp.
\end{eqnarray}
The corresponding background is $ZZZ$, where the $Z$ also decays to lepton pair.
 However, the cross section of the signal is so small that it might hardly be detected
 in the future. Thus, we do not show the relevant numerical results in Table III.

\vspace{0.5cm} \noindent{\bf 4. CONCLUSIONS AND DISCUSSIONS }

The model \cite{12046599} which is studied in this paper can explain the smallness of the neutrino masses.
The empirical masses of the neutrinos $m_\nu\sim 10^{-1}$ eV can be achieved by Majorana quintuplets
$\Sigma_R$ and scalar quadruplet $\Phi$ which transform as (1,5,0) and (1,4,-1) under the SM gauge group, respectively.
The quintuplet heavy leptons can couple to the SM particles, the couplings are proportional to the
mixing matrix $V_{l\Sigma}$ between the heavy leptons and the SM leptons.
The LHC can provide enough energy and high luminosity to produce such heavy leptons and
detect their signatures.

In this paper, we investigated pair production and the associated production of the heavy leptons and found that
they are copiously produced by the quark-antiquark annihilation mediated by the neutral and charged SM gauge bosons. Considering the multiple decay modes of the heavy leptons and the SM gauge bosons, we studied several types of signals with different BRs.
Firstly, we carried out a full simulation for the signals $2l^\pm l^\mp 2j \rlap/E_T$, $2l^\pm 2l^\mp 2j$ and
the relevant SM backgrounds. The results revealed that the two signals have a large statistical significance.
Furthermore, we also studied the LNV signals $3l^\pm l^\mp 2j$ and $3l^\pm 2l^\mp \rlap/E_T$.
The cross sections of the backgrounds are smaller than those of the LNV signals, thus, we only applied the basic cuts
on the signal and background events. For the $3l^\pm 3l^\mp$ signal, there were few signal events with
high integrated luminosity. Based on the above results, the possible signatures of the heavy leptons could be
detected at the 14TeV LHC in the near future.

These production channels could also provide a lepton flavor violating signal at the LHC,
where the two leptons from the heavy leptons decay are a electron and a muon, and the $W$ and the $Z$ from
the heavy lepton decays hadronically so that the signal has a large BR. The signal can be
$e^- \mu^+4j$ and the SM background will be dominated by $W^+W^-4j$.
The important difference between signal and background is that the background contains missing energy in the form of neutrinos.
Detailed studies, including sample selection and standard cuts, will be presented in the future.

Reference \cite{hantao} has studied the phenomenology of a lepton triplet in both low energy experiments and at the LHC.
There are some differences between the triplet leptons and the quintuplet leptons that we studied here.
First of all, Ref. \cite{hantao} predicted the heavy leptons in the form of a vector-like triplet with hypercharge $\pm 1$.
Three generations of fermion quintuplets with zero hypercharge were predicted in the model here and
they only have only a right-hand component. Second, the heavy leptons have different couplings with gauge bosons
in the two cases. Thus, they have different production cross sections at the LHC. With the different BRs,
we can get different signal rates. Third, the quintuplet leptons can produce the LNV
signals $3l^\pm l^\mp 2j$ and $3l^\pm 2l^\mp \rlap/E_T$ which are the key features of the model. In addition, we applied different basic
cuts and chose different simulation methods according to the kinematical differences between the signals and the backgrounds
to extrude the signals and suppress the backgrounds in our simulation.
 \vspace{4mm}
\\
\vspace{4mm} \textbf{Acknowledgments}\\
This work was supported in part by the National Natural Science Foundation of
China under Grants Nos.11275088, 11175251, 11205023,  the Natural Science Foundation of the Liaoning Scientific Committee (No. 2014020151) and Liaoning Excellent Talents in University (LJQ2014135).

\end{document}